\documentclass{PoS}

\title{TASI Lectures on Applications of Gauge/Gravity Duality}

\ShortTitle{Applications of Gauge/Gravity Duality}

\author{\speaker{Oliver DeWolfe}
\\
      Department of Physics, 390 UCB, University of Colorado, Boulder, CO 80309, USA \\
Center for Theory of Quantum Matter, University of Colorado, Boulder, CO 80309, USA\\
      E-mail: \email{oliver.dewolfe@colorado.edu}}

\abstract{We introduce the gauge/gravity, or AdS/CFT, correspondence with an eye towards its application to strongly coupled systems. We provide an overview of the duality, including the AdS/CFT dictionary and holographic renormalization. We then discuss simple correlation functions in the context of renormalization group flow geometries,  holographic thermodynamics and an application to the phase diagram of QCD, real-time correlators and the shear viscosity of strongly coupled field theories, holographic superconductors, and the application of holographic fermionic response to strange metals. These lectures were delivered at TASI 2017.}

\FullConference{Theoretical Advanced Study Institute Summer School 2017 "Physics at the Fundamental Frontier"\\
		 4 June - 1 July 2017\\
		 Boulder, Colorado \flushright{COLO-HEP-596}}

\begin{document}

\section{Introduction}

Dualities are among the most fascinating, and the most useful, phenomena of physics. A duality expresses that the same physical system can be described in more than one way, with the two ``dual" descriptions consisting of entirely different degrees of freedom, interacting in different ways. Often, one dual description is strongly coupled while the other is weakly coupled. We have the intuition that near weak coupling we have an approximately free set of degrees of freedom, but as the coupling increases those degrees of freedom can bind to one another in such a way that they are no longer independent, and treating the problem in the original variables is no longer easy. Sometimes, however, it occurs that the bound states of the original degrees of freedom behave as a new set of variables that are themselves weakly coupled with respect to a different set of collective interactions, even though the original interactions are strong. Such a situation can enable physicists to run back to their beloved weakly-coupled perturbation theory in situations where it originally seemed wildly inappropriate, by switching to the new degrees of freedom.

The gauge/gravity, or AdS/CFT, correspondence \cite{Maldacena:1997re, Gubser:1998bc, Witten:1998qj} is a particularly remarkable duality. The two dual sets of degrees of freedom and their interactions are profoundly different. On one side we have a theory of (in general quantum) gravity in a higher spacetime dimension, while in the other case we have a theory without gravity, in a lower spacetime dimension; the original and most famous realization is of a duality between string theory in an anti-de Sitter (AdS) space, and a conformal field theory (CFT) in one dimension less. One might think that a duality between a gravitational theory and a non-gravitational theory should not occur, and one might also think that a duality between physics in different spacetime dimensions {\em really} should not occur. Yet the gauge/gravity correspondence does both, and in fact, it is essential that it is both together: this correspondence provides a realization of the {\em holographic principle} \cite{tHooft:1993dmi, Susskind:1994vu, Bousso:2002ju}, which holds that any theory of quantum gravity is in some sense massively redundant, and should be describable in terms of fewer degrees of freedom living on some sort of ``boundary" of spacetime. Thus one member of the dual pair having gravity, and the same member living in a higher-dimensional space, are two sides of the same coin, and one is necessary for the other.

The gauge/gravity correspondence is interesting in both directions. Starting from the non-gravitational field theory side, one has in principle a definition of quantum gravity in certain backgrounds, and can attempt to understand the emergence of spacetime from the CFT data. This program has many exciting developments and is described in the TASI 2017 summer school by lectures from Harlow \cite{Harlow:2018fse} and from Headrick; see also the TASI 2015 lecture notes from van Raamsdonk \cite{VanRaamsdonk:2016exw}.
The other way to proceed is to attempt to use what we know about gravity to learn about non-gravitational field theory. In particular, the realization of the strong/weak coupling duality dichotomy in the case of the AdS/CFT correspondence is that when the field theory has a large number of degrees of freedom (``large N") and is strongly coupled, the dynamics on the gravity side reduce to classical general relativity interacting with other fields. Thus we can ignore the (admittedly fascinating) story of quantum gravity, and attempt to use what we know about classical gravity to learn things about strongly coupled field theories. This is the subject of this review. Such applications were also explored in the last two lectures of Erdmenger's AdS/CFT lectures at TASI 2017. For other reviews of such applications, see
the monumental review by Hartnoll, Lucas and Sachdev \cite{Hartnoll:2016apf}, 
McGreevy's lectures from TASI 2015 \cite{McGreevy:2016myw}, lectures on AdS/CFT and heavy ions like \cite{CasalderreySolana:2011us, Adams:2012th, DeWolfe:2013cua} and others on applications to condensed matter such as \cite{McGreevy:2009xe, Hartnoll:2009sz}.

There have been a number of useful reviews of AdS/CFT itself. The ``MAGOO" Physics Report \cite{Aharony:1999ti} is a twentieth-century classic, including background on the large-N limit and a discussion of AdS$_3$, among many other things. Klebanov's 1999 TASI lectures \cite{Klebanov:2000me} provide a lot of background on the string/brane systems that led to AdS/CFT, and a nice description of the remarkable holographic system arising from branes at a conifold singularity.
D'Hoker and Freedman's lectures at TASI 2001 \cite{DHoker:2002nbb} provide encyclopedic detail of supersymmetry in AdS/CFT and higher-point correlation functions, among other things.
Polchinski's 2010 TASI lectures \cite{Polchinski:2010hw} contain an alternate argument not involving strings or branes for the existence of a gravity dual starting with a field theory, and 
Penedones's 2015 TASI lectures \cite{Penedones:2016voo} take the viewpoint of starting with physics in AdS space and deriving that something looking  like a dual CFT must exist.

With all of these reviews out there, I will not attempt to be encyclopedic here. Instead I will try to provide an entry point into the subject, discussing some important basics and classic examples. Maybe a better name for these lectures would have been ``Gauge/gravity duality for applications," for I will focus on the gravity calculations and the nature of the duality, discussing the applications (in particular the strong nuclear interaction of quantum chromodynamics, and the theory of high-temperature superconductors and the associated strange metals) in a more cursory fashion, as motivation for the gravity calculations.  Hopefully the reader will finish these lectures with a solid foundation in the subject, as well as feeling intrigued and ready to delve more deeply into the literature.

\section{Gauge/Gravity duality}

``The AdS/CFT correspondence" is not the best name for such a profound and far-reaching duality. Besides being stuffed with two initially-mysterious acronyms, it also suffers from being overly specific: the correspondence can easily be generalized to cases where the gravity theory does not live in anti-de Sitter space, and the dual theory is not a conformal field theory. Nonetheless the name has stuck, and we will use it like everyone does. ``The gauge/gravity correspondence" is a little better, since it is more general, and we always have a gravity theory on one side, and in all our examples where the other side is known it will be a gauge theory, so we will use this name as well, mostly interchangeably. 

In this section we will provide an overview of the correspondence, as well as introducing anti-de Sitter space and considering the limits of validity of classical gravity, while mentioning a few famous examples of the duality.

\subsection{Overview of the correspondence}

Let us begin by providing an overview of how the AdS/CFT correspondence works. All of our examples will have the gravity theory living in a spacetime that is at least asymptotically anti-de Sitter, so we will assume that in what follows, though there are generalizations.

The essential features of the duality are:

\begin{itemize}
	{\item On one side of the correspondence we have a theory of gravity, with a metric and other fields. On the other side is a gauge theory. }
	
	{\item The same symmetries act on both sides of the correspondence. In particular the isometries of the gravity theory geometry are exactly the spacetime symmetry group of the field theory. In this way a change of scale in the field theory is associated with motion through the extra dimension on the gravity side. Additionally, gauge groups of the gravity theory match global symmetries of the field theory.  The field theory's gauge group is an exception; it does not appear on the gravity side.}
	
	{\item The AdS/CFT dictionary associates to each field in the gravity theory, a gauge-invariant operator in the gauge theory, with the same symmetry properties.}
	{\item Doing physics in anti-de Sitter space requires not just initial conditions, but also boundary conditions at spatial infinity (``the boundary") to be specified. Setting these boundary conditions for gravity theory fields is mapped by the correspondence to adding sources or turning on expectation values for the dual operators. From this, correlation functions of gauge theory operators can be calculated by determining the gravity theory's response to changing boundary conditions. Intuitively, we say the gauge theory ``lives" at the boundary of the gravity theory.}
	{\item Black hole thermodynamics in the gravity theory is mapped to regular thermodynamics in the gauge theory: a black hole in AdS space with a certain Hawking temperature, entropy etc. corresponds to a state in the field theory with the same thermodynamic properties.}
	
\end{itemize}

One of the most important features in the correspondence is the particular nature of anti-de Sitter space, its boundary at infinity and the need for boundary conditions there, so let us introduce this geometry.

\subsection{Anti-de Sitter and asymptotically anti-de Sitter spaces}

Anti de-sitter space is the maximally symmetric space with Lorentzian signature and negative curvature. It is notable for the importance of its ``boundary", the limit of the geometry at large spatial distance: unlike for example Minkowksi space, where the boundary is infinitely far away and doesn't bother us, in AdS space the boundary may be reached by a signal which then returns in finite proper time. This makes the boundary act like a real place,  and consequently we have to understand what's going on there.

 We will discuss anti-de Sitter space in $D \equiv d+1$ spacetime dimensions. There are many sets of coordinates that can be used to describe some or all of the space. The full (geodesically complete) geometry is called {\em global} AdS, and has a boundary with topology $R \times S^{d-1}$. For our applications, we will instead be interested in a subset of AdS space, the so-called {\em Poincar\'e patch}, whose metric can be written as
\begin{eqnarray}
\label{AdSMetric}
ds^2 = {r^2 \over L^2} (-dt^2 + d\vec{x}^2) + {L^2 dr^2 \over r^2} \,,
\end{eqnarray}
where there are $d-1$ spatial coordinates $\vec{x}$, one time coordinate $t$ and the spatial ``radial coordinate" $r$, for a total of $d+1$ dimensions.\footnote{Another useful coordinate system uses the coordinate $z \equiv L^2/r$, where the metric looks like $ds^2 = {L^2 \over z^2} (-dt^2 + d\vec{x}^2 + dz^2)$ and the boundary is at $z=0$.} Here $L$ is the characteristic length scale of the geometry, called the AdS radius. The constant negative curvature is expressed by the Ricci scalar,
\begin{eqnarray}
R = - {d(d+1)\over L^2} \,.
\end{eqnarray}
Anti de-Sitter space is a solution to Einstein's equations with a cosmological constant,
\begin{eqnarray}
R_{\mu\nu} + {1 \over 2}R g_{\mu\nu}  +\Lambda g_{\mu\nu} = 0\,, \quad \quad \Lambda = - {d (d-1) \over 2 L^2} \,. 
\end{eqnarray}
Often this $D$-dimensional Einstein equation arises as the dimensional reduction of a higher dimensional theory dimensionally reduced on a positive-curvature geometry like a sphere: in the higher dimension we will have $AdS_D \times S^q$, with a $D$-form or $q$-form field strength $F_D$ or $F_q$ (generalizations of the two-form field strength of electromagnetism) on the AdS or compact factor, respectively, inducing the cosmological constant when reduced to the lower-dimensional theory.

The boundary of the Poincar\'e patch of anti-de Sitter space is at $r \to \infty$, and has topology $R^d$; one can see that the  induced metric on a slice at fixed $r$ is just an overall constant times the $d$-dimensional Minkowski metric. A null ray may leave $r=0$, reach $r=\infty$ and return in finite proper time for a timelike observer at the origin. As a result, to understand physics in anti-de Sitter space it is not enough to specify initial conditions on a spacelike hypersurface like $t=0$; we must also set boundary conditions at infinity. Thus even for fixed bulk dynamics, each set of boundary conditions defines a different physical theory living in the space.

In the AdS/CFT correspondence, these boundary conditions are intimately related to the connection between the gravity theory and the dual field theory. The dual field theory lives on a space with the topology of the AdS boundary, and it is often convenient to think of it as ``living on the boundary". For this reason, we will define the correspondence more precisely before we impose the AdS boundary conditions, so we can discuss what we are doing on both sides of the duality simultaneously.

Once we put things inside the geometry, the metric will be deformed and it will not be precisely AdS space anymore. Sometimes we will want to put some pretty big things in there, like black holes. However, such deformations will still leave the geometry asymptotically approaching AdS at infinity. Thus we will be interested in asymptotically AdS geometries of the form
\begin{eqnarray}
\label{AsymptoticAdS}
ds^2 = e^{2A(r)} (- h(r) dt^2 + d\vec{x}^2) + e^{2B(r)} dr^2 \,,
\end{eqnarray}
along with other fields depending on $r$, reducing to the AdS metric (\ref{AdSMetric}) as $r \to \infty$. This breaks a certain set of symmetries of the space; as we discuss momentarily, this must correspond to breaking the analogous symmetries on the field theory side. It is possible to consider even less symmetric geometries, for example breaking spatial homogeneity; such examples are quite interesting, but we will not have time to consider them here.

\subsection{Validity of classical gravity, large $N$ and some famous cases}

A theory of gravity always has an associated Newton's constant $G_N$, which we may define as the coefficient of the Einstein term in the action: $S = (1 / 16 \pi G_N) \int d^Dx \, R$. From $G_N$ we can define the Planck length as $G_N = \ell_P^{D-2}$,  characterizing the length scale at which quantum gravity effects should appear. String theory has another length scale, the string scale $\ell_s$, characterizing the size of the string. 

The strongest form of the gauge/gravity duality asserts that the full quantum theory of gravity on a particular spacetime is dual to the appropriate field theory, for all values of their respective parameters. We, however, will stay within the classical gravity limit, which says that the AdS radius $L$, which is the characteristic length of the geometry, is much bigger than the Planck length $\ell_P$ and the string length $\ell_s$:
\begin{eqnarray}
\label{Limits}
L \gg \ell_P, \ell_s \,.
\end{eqnarray}
Thus the quantum or stringy nature of the geometry is not visible on the characteristic scale of the curvature.

In the dual field theory, this corresponds to the limit of a large number of degrees of freedom (large $N$) and strong coupling. In the case of the gauge theories we study, the large number of degrees of freedom means a large number of colors for the gauge group (for example, $SU(N)$ gauge theory with large N). Studying this large N limit as a way to simplify theories like QCD predates the AdS/CFT correspondence by two decades (for a brief review see \cite{Aharony:1999ti, McGreevy:2009xe}, or for more detail Coleman's lectures ``1/N" in \cite{Coleman}). To avoid divergent loop diagrams with an infinite number of fields running in the loop one must simultaneously tune the coupling  to zero, while keeping an effective coupling that is a product of the vanishing original coupling and the diverging number of colors, called the 't Hooft coupling, fixed. The limit (\ref{Limits}) corresponds to the large $N$, large 't Hooft coupling limit of the dual field theory.

To be a little more precise, here are two fundamental examples of the AdS/CFT correspondence, both discovered in \cite{Maldacena:1997re}:

\begin{itemize}
	{\item Type IIB string theory has, among other fields, a metric and a four-form gauge field with a self-dual five-form field strength. This theory on $AdS_5 \times S^5$ with $N$ units of five-form flux on each five-dimensional spacetime factor is dual to ${\cal N}=4$ Super-Yang-Mills theory in four spacetime dimensions with gauge group $SU(N)$, a particular non-abelian gauge theory cousin of QCD consisting of gauge fields, 4 fermions and 6 scalars all in the adjoint representation of $SU(N)$, that is also an exactly superconformal quantum field theory. The limit (\ref{Limits}) reduces us to Type IIB supergravity, and corresponds in the field theory to $N \to \infty$ with the 't Hooft coupling $\lambda \equiv g^2_{\rm YM} N$ fixed and large.}
	
	{\item Eleven dimensional M-theory has a metric and a three-form gauge potential with a four-form field strength. This theory on $AdS_4 \times S^7$ with $N$ units of four-form flux on $AdS_4$ corresponds to ABJM theory \cite{Aharony:2008ug}, a superconformal Chern-Simons-matter theory in three dimensions with gauge groups $U(N)_k \times U(N)_{-k}$, where $k$ is the Chern-Simons level. The limits (\ref{Limits}) again correspond to $N \to \infty$, $\lambda$ fixed and large where the 't Hooft coupling is now $\lambda \equiv N/k$.}
 	
\end{itemize}
We will discuss these in a little more detail in section~\ref{NScalingSec}.

Studying these very special, highly-symmetric superconformal theories in their strongly coupled limits can be interesting in its own right. ${\cal N}=4$ Super-Yang-Mills has been called ``the simple harmonic oscillator of the $21^{st}$ century" for its iconic role as a prototype gauge theory and quantum field theory that can be studied from many angles and generalized in many directions, while ABJM theory appears to play no less of a fundamental role in the pantheon of quantum field theories as the avatar of three-dimensional Chern-Simons-matter theories, and is less famous likely just due to our (admittedly understandable) preference for four-dimensional physics.\footnote{The final member of the triumvirate of highly special superconformal theories is the six-dimensional $(2,0)$ theory, which also has a gravity dual, and since the others can be obtained from it under dimensional reduction it may be the most special of all. But because it is the least-well understood from the field theory perspective, we will not use it as an example.} But we are also motivated by less symmetric strongly coupled systems:

\begin{enumerate}
	{\item Quantum chromodynamics (QCD) is the theory of the strong nuclear force, but cannot be studied at low energies in perturbation theory as quantum electrodynamics (QED) can be, due to its strong coupling. The large-$N$ limit in gauge theories was developed in an attempt to make QCD tractable. Anything we can learn about its strong coupling behavior from AdS/CFT would be highly desirable.}
	
	{\item High-$T_c$ superconductors are strongly correlated electron materials that arise in the study of condensed matter systems. They are effectively two-spatial-dimensional, and their phase diagrams include a so-called ``strange metal" phase that appears not to be describable in terms of an effective quasiparticle description. Again, 
	a holographic description would be most useful.} 
\end{enumerate}
In what follows, we will use these systems as motivation and examples for what kind of physical systems the gauge/gravity correspondence can make contact with.

\section{The AdS/CFT dictionary}

The AdS/CFT correspondence connects a gravity theory living in (asymptotically) $AdS_{d+1}$ to a field theory living on $R^d$. The first part of the matching between the two theories is the matching of symmetries. After describing this we will establish the holographic dictionary between gravity theory fields and gauge theory operators, give the mathematical statement of the correspondence, and explore aspects of it in some detail for a single scalar field.

\subsection{Symmetries}

\medskip\noindent
\underline{Spacetime symmetries}

A Lorentz-invariant field theory living on $R^d$ is symmetric with respect to the $d$-dimensional Poincar\'e group, including translations, rotations and Lorentz boosts. When the theory is also conformal, the spacetime symmetry group is enhanced to the conformal group $SO(d,2)$, which includes the symmetries above along with the scale transformation,
\begin{eqnarray}
D: t \to \lambda t \,, \quad \vec{x} \to \lambda \vec{x}\,,
\end{eqnarray}
and the $d$ special conformal transformations, which we won't have need to discuss. Conformal field theories describe physics invariant under this overall scale transformation, which means they lack any length or mass scale. The ultraviolet (high-energy) and infrared (low-energy) limits of a general quantum field theory are CFTs, and as such CFTs represent fundamental building blocks for the study of quantum field theory, as well as being relevant to physical systems from critical phenomena to string theory.

$AdS_{d+1}$ is a higher-dimensional space, yet its isometry group --- the group of coordinate transformations that preserve the metric --- is also precisely $SO(d,2)$. The translations, rotations and Lorentz boosts on the coordinates $t, \vec{x}$ exactly match those acting on the analogous coordinates in the field theory; it is this matching of symmetries that really lets us identify the space the field theory lives on with a constant-$r$ slice of the gravity theory. Furthermore, the AdS analog of the scale transformation is
\begin{eqnarray}
\label{GravityScale}
D: t \to \lambda t \,, \quad \vec{x} \to \lambda \vec{x}\,, \quad r \to {r \over \lambda} \,.
\end{eqnarray}
This acts as a scale transformation on the $t, \vec{x}$ coordinates, while also moving us in the radial direction. This identification leads us to one of the profound aspects of AdS/CFT,
\begin{center}
	{\em Moving in the radial direction of the gravity theory corresponds to a change of scale in the field theory.}
\end{center}
As we shall see, {\em breaking} this symmetry on one side corresponds to breaking it on the other side, as well. Thus asymptotically AdS geometries that have something sitting at some value of $r$, or otherwise deviate from pure AdS as we move in the radial coordinate, correspond to a field theory where scale invariance is broken, and moving through the radial direction corresponds to moving through the scales of the theory.

\medskip\noindent
\underline{Global symmetries}

A global symmetry $G$ of the field theory is realized by a {\em gauge} symmetry $G$ in the gravity theory: each global conserved current in the field theory $J$ is associated to a fluctuating gauge field $A$ on the gravity side. 

For an example of this, ${\cal N}=4$ Super-Yang-Mills has an $SO(6)$ global symmetry under which the gauge field is invariant, the fermions transform in the ${\bf 4}$ and the scalars in the ${\bf 6}$. The gravity dual is type IIB string theory on $AdS_5 \times S^5$, and in reducing from 10 to 5 dimensions, the spacetime metric modes with one index on the $S^5$ reduce to $SO(6)$ gauge fields, realizing the $SO(6)$ isometry group of the compact $S^5$ factor. This $SO(6)$ gauge symmetry of the gravity theory matches the $SO(6)$ global symmetry of the field theory. Similarly, ABJM theory has an $SO(8)$ symmetry that is matched to the $SO(8)$ isometry group of the compact $S^7$ in $AdS_4 \times S^7$ on the gravity side.

\medskip\noindent
\underline{Fermionic symmetries}

Fermionic symmetries (``supersymmetries") must also match between the field theory and gravity sides. For ${\cal N}=4$ Super-Yang-Mills, the ${\cal N}=4$ four-dimensional spinor supercurrents give us 16 supercharges worth of fermionic symmetries; however, the exact conformal invariance of the theory leads to 16 more ``superconformal" fermionic symmetries. These 32 supercharges combine with the conformal group $SO(4,2) \simeq SU(2,2)$ and the global symmetry group\footnote{The fact that $SO(6)$ does not commute with the supercharges makes it an {\em R-symmetry group}. R-symmetries commute with the spacetime symmetries, but both bosonic groups do not commute with supersymmetry, joining them all in one larger (super)group.} $SO(6) \sim SU(4)$ to make a supergroup called $SU(2,2|4)$. Meanwhile on the gravity side, type IIB string theory on $AdS_5 \times S^5$ also preserves 32 supercharges (the maximal amount of supersymmetry in ten dimensions) and again the overall group of symmetries is the supergroup $SU(2,2|4)$. We won't have anything to say about supergroups, but it is an essential check of the correspondence that they too match on both sides.

\medskip\noindent
\underline{Gauge symmetries}

The field theory in general will also have a gauge symmetry --- the ``gauge" part in ``gauge/gravity correspondence". In the case of ${\cal N}=4$ Super-Yang-Mills, this is $SU(N)$. However, there is {\em no} appearance of this gauge group on the gravity side, except for the presence of the parameter $N$ as the total five-form flux.

Is this is a problem for the correspondence? It is not. A gauge symmetry, fundamentally, is not a true symmetry at all, even though we use the language of symmetries to describe it. Instead it is a redundancy of description; physical quantities must be gauge-invariant. When one has a duality, it is not necessary that the gauge symmetries match on both sides of the duality, since the different gauge symmetries are characteristic of the different degrees of freedom used. Only the gauge-invariant variables must match across the duality. This is the case with AdS/CFT: only gauge-invariant operators in the field theory will match with fields on the gravity side.

\subsection{Field/operator correspondence}

Having described the matching between the symmetries of the two sides of the correspondence, we next turn to the connection between the variables.  A fundamental part of the AdS/CFT dictionary is that there is an association between each {\rm field} $\phi(r, \vec{x}, t)$ of the gravity theory and a {\em gauge-invariant operator} ${\cal O}(\vec{x}, t)$ of the field theory:
\begin{eqnarray}
\label{Dictionary}
\phi(r, \vec{x}, t) \quad \leftrightarrow \quad {\cal O}(\vec{x}, t) \,.
\end{eqnarray}
Consider for concreteness a scalar field $\phi(r, \vec{x}, t)$ on the gravity side; we will say more about other spin fields later. Assume its quadratic action takes the Klein-Gordon form,
\begin{eqnarray}
	S_{\rm KG} = {1 \over 2 \kappa^2}  \int d^{d+1}x \sqrt{-g} \left( - {1 \over 2}(\partial \phi)^2 - {1 \over 2}m^2 \phi^2 \right) \,,
\end{eqnarray}
where we have included an overall factor $1/2\kappa^2$ with mass dimension $d-1$; in many supergravity theories all terms share the same overall normalization with the Einstein-Hilbert action, which renders the scalars dimensionless. Newton's constant $G_N$ of the gravity theory is then related by $\kappa^2 = 8 \pi G_N$. This normalization will not affect the equations of motion, but the overall value of the action will be important, as we will see.

The corresponding Klein-Gordon equation of motion is
\begin{eqnarray}
\label{KGEqn}
\left( - {1 \over\sqrt{-g}} \partial_\mu \sqrt{-g} g^{\mu\nu} \partial_\nu + m^2 \right) \phi = 0 \,.
\end{eqnarray}
Near the boundary $r \to \infty$ the geometry approaches AdS space. The solution to (\ref{KGEqn}) then approaches
\begin{eqnarray}
\label{KGSoln}
\phi(r \to \infty, \vec{x}, t) = {\alpha(\vec{x},t) L^{2\Delta_-} \over r^{\Delta_-}} + \cdots + {\beta(\vec{x},t)L^{2\Delta_+} \over r^{\Delta_+}} + \cdots \,,
\end{eqnarray}
where we defined the exponents
\begin{eqnarray}
\Delta_\pm \equiv {d \over 2} \pm \sqrt{\left(d \over 2\right)^2 + m^2 L^2} \,.
\end{eqnarray}
Since (\ref{KGEqn}) is a second-order differential equation, it has two independent solutions on the boundary, represented by the leading $\alpha(\vec{x}, t)$ and the subleading $\beta(\vec{x}, t)$. (Note that there may be other terms bigger than the $\beta(\vec{x},t)$ term, as indicated by the dots in the middle, but these all depend on $\alpha(\vec{x},t)$ and vanish in the $\alpha \to 0$ limit; $\beta$ is the leading independent term.) Since $\phi(r, \vec{x}, t)$ is a coordinate scalar, the scaling isometry (\ref{GravityScale}) indicates that $\alpha(\vec{x}, t)$ and $\beta(\vec{x}, t)$ must scale in such a way so as to cancel the transformations of $r^{\Delta_-}$ and $r^{\Delta_+}$; this implies that they behave as $d$-dimensional objects with dimensions $\Delta_-$ and $\Delta_+$, respectively. We inserted the factors of $L$  into (\ref{KGSoln}) so these are their engineering dimensions, as well.\footnote{Using the variable $z\equiv L^2/r$ absorbs all the factors of $L$, and this can be calculationally more convenient. We mostly stick with the variable $r$ as it emphasizes that the boundary lies at infinite distance, and it is more commonly used in the solutions we will encounter.}

We must now impose boundary conditions at $r \to \infty$ to make our AdS theory well-defined.  Constraining half the degrees of freedom at the boundary is sufficient. The simplest choice is to remove the leading term, $\alpha(\vec{x}, t) = 0$. More generally, we can constrain $\alpha(\vec{x}, t)$ to take specified values:
\begin{eqnarray}
\label{RegularQuant}
\alpha(\vec{x}, t) = J(\vec{x}, t) \,,
\end{eqnarray}
where $J(\vec{x}, t)$ is chosen by us and fixed. The other near-boundary solution, $\beta(\vec{x}, t)$ is unspecified and allowed to fluctuate dynamically. Then, initial conditions on a spacelike hypersurface $\Sigma$ plus these boundary conditions lead to a unique time evolution for the field throughout spacetime.

When the boundary conditions (\ref{RegularQuant}) are imposed, we refer to this as the {\em regular quantization}. We will associate the fluctuating mode $\beta(\vec{x}, t)$ with the dual field theory operator ${\cal O}$, in a way which will become more precise momentarily, and thus the dimension of ${\cal O}$ is the dimension of $\beta$,
\begin{eqnarray}
\label{RegularDim}
\Delta_{{\cal O}} = \Delta_+  = {d \over 2} + \sqrt{\left( d \over 2\right)^2 + m^2 L^2}\,.
\end{eqnarray}
We are now prepared to make the statement of the gauge/gravity correspondence. It can be viewed as an equality of path integrals, where the gravity path integral with boundary conditions (\ref{RegularQuant}) on fields is equal to the field theory path integral with the same functions $J(\vec{x},t)$ turned on as {\em sources} for the operators \cite{Gubser:1998bc, Witten:1998qj}:
\begin{eqnarray}
\label{AdSCFT}
Z_{\rm grav}[\phi; \alpha(\vec{x},t) = J(\vec{x},t)] = Z_{\rm CFT}[{\rm source \ for} \ {\cal O}(\vec{x},t) \ {\rm  is} \ J(\vec{x},t) ] \,.
\end{eqnarray}
Of course, there is a difficulty in general with characterizing the left-hand-side of this equation at all. For arbitrary $N$ and $\lambda$ it is a quantum gravity theory which we are not sure how to write down.\footnote{In fact, away from the classical gravity limit, a reasonable way to proceed is to {\em define} the left-hand-side as whatever it needs to be to satisfy this equation. We are only comfortable doing this because of the nontrivial checks on the correspondence that can be carried out in the cases where we can characterize the left-hand-side.}
In these lectures, however, we are working in the large-$N$, large 't Hooft coupling limit. In that case, the left-hand-side reduces to a saddle point, localizing the fields to solutions to the classical gravitational equations of motion. We then have
\begin{eqnarray}
\label{AdSCFT2}
\exp{i S_{\rm grav}[\phi; \alpha = J]} = \left\langle \exp{i \int d^dx \; J(\vec{x}, t) {\cal O}(\vec{x}, t)} \right\rangle_{\rm CFT} \,. 
\end{eqnarray}
Now the left-hand-side is to be understood as the classical gravitational action, with both quantum and stringy corrections neglected, evaluated on solutions of the classical equations of motion. The right-hand-side is unchanged from the previous expression, and just written in a different way.

Let us say again what we have done in words: each gravity field is associated to a field theory operator. Imposing boundary conditions for the gravity fields corresponds to turning on sources for the corresponding operators. The field theory path integral with these sources is equal to the exponential of the gravity action with the corresponding boundary conditions.

A few remarks:

\begin{itemize}
	{\item The equations (\ref{AdSCFT}) and (\ref{AdSCFT2}) are written as if there is only one scalar field $\phi$ and one spinless dual operator ${\cal O}$, but in general there will be any number of fields of different spins. Each field will have a corresponding boundary condition corresponding to turning on a source for the dual operator. We will discuss the other spins more later, but for now will stick to the example of a scalar field $\phi$.}
	{\item Since $\alpha(\vec{x},t)$ has dimension $\Delta_- = d - \Delta_+$ as far as the coordinates $\vec{x}$ and $t$ are concerned, so must $J(\vec{x},t)$, and thus it has the correct dimension to be a source for the $\Delta_+$-dimensional operator ${\cal O}(\vec{x}, t)$ in $d$ dimensions.} 
	{\item One can pass from a Lorentzian formulation to a Euclidean one by replacing the $i$ factors in (\ref{AdSCFT}) and (\ref{AdSCFT2}) with minus signs.}
\end{itemize}

We now flesh out this correspondence by studying a few important details. 

\subsection{Relevant operators and the Breitenlohner-Freedman bound}

Looking at the formula (\ref{RegularDim}) for the operator dimension, we see that requiring $m^2 \geq 0$ gives us dual operators of dimension $\Delta \geq d$: that is, irrelevant and marginal operators. What about relevant operators?

In quantum field theory in Minkowski space, we are used to requiring $m^2 \geq 0$ for a scalar field in a stable vacuum. If $m^2 < 0$, unstable modes will exist, signaling that we are not in the correct vacuum. In anti-de Sitter space, however, the story is a little different. As shown by Breitenlohner and Freedman, it is possible to have $m^2 < 0$ without leading to an instability. The total energy receives negative contributions from the mass term, but receives compensating positive contributions from the kinetic energy term. As a result, the net energy can be positive, with no instabilities \cite{Breitenlohner:1982jf, Breitenlohner:1982bm}.

However, this is only possible if the mass-squared is not too negative. In order to have a perturbatively stable vacuum, scalar masses must satisfy the {\em Breitenlohner-Freedman (BF) bound}, 
\begin{eqnarray}
m^2 L^2 \geq - {d^2 \over 4} \,.
\end{eqnarray}
The dimension formula (\ref{RegularDim}) continues to hold for allowed negative-mass cases, and we see that scalars with mass-squared going down to the BF bound get us dual operators with dimensions going down to $\Delta = d/2$; we now have found some of the relevant operators. (There is a way to take the dimension down even further, to $\Delta = d/2 - 1$, as we will see momentarily.)

The case where the mass-squared precisely saturates the BF bound $m^2 L^2 = -d^2/4$ is special, because here $\Delta_+ = \Delta_-$ and the asymptotic solution (\ref{KGSoln}) no longer holds. Instead we find
\begin{eqnarray}
\label{BFBoundSaturating}
\phi(r \to \infty, \vec{x}, t) = {\alpha(\vec{x},t)  L^{d/2} \log r \over r^{d/2}} +  {\beta(\vec{x},t) L^{d/2}\over r^{d/2}} + \cdots \,,
\end{eqnarray}
where a logarithm has showed up to distinguish the two independent solutions. We can take the boundary condition $\alpha(\vec{x},t) = J(\vec{x},t)$ to correspond to a source for a dimension-$d/2$ operator, as before.

\subsection{Holographic renormalization and one-point functions}

The gravity side of the correspondence involves the gravitational action $S_{\rm grav}$, evaluated on solutions to the equations of motion. Unfortunately, this turns out not to be finite! Let's see how this goes for our example of a single scalar field $\phi$, where $S_{\rm grav} = S_{\rm KG}$. We'll begin by regulating things, cutting off the spacetime at a large value $r=R$. Integrating $S_{\rm KG}$ by parts, we find
\begin{eqnarray}
\label{RegulatedKG}
S_{\rm KG} = {1 \over 4\kappa^2 } \int d^{d+1}x \sqrt{-g} \phi \left( \square - m^2\right) \phi - {1 \over 4\kappa^2} \int_{r=R} d^dx \sqrt{-h}\, \phi n^\mu \partial_\mu \phi \,,
\end{eqnarray}
where the second term is a boundary term evaluated at $r=R$, with $h_{\mu\nu}$ the induced boundary metric and $n^\mu$ a normal vector field. The bulk term vanishes on solutions to the equations of motion, but plugging in the asymptotic $\phi$ solution (\ref{KGSoln}) and using $\sqrt{-h} \sim r^d$, $n^r \sim r$ in $AdS_{d+1}$, we find that the boundary term diverges as we take $R \to \infty$.

Thus it seems like our impressive fundamental gauge/gravity relation involves a left-hand-side that's infinite! But that's okay; the right-hand-side is infinite too. We are dealing with a quantum field theory after all, and quantum field theories have ultraviolet divergences if we take them to be valid to arbitrarily high energy scales (arbitrarily short distances).

So to make sense of AdS/CFT, we need to do what we always do with quantum field theory: we need to regulate it to render the divergences finite, and then renormalize by adding suitable counterterms so that we get finite results when the regulator is removed. Now we're just going to carry out this regularization and renormalization on the gravity side. The scale transformation (\ref{GravityScale}) tells us that large distances in gravity ($r \to \infty$) match up with short distances in field theory ($t, \vec{x} \to 0$) and thus it is natural that the short-distance, ultraviolet divergences of QFT are showing up as the long-distance divergences in AdS gravity. Cutting off the geometry at $r=R$ as in (\ref{RegulatedKG}) is precisely an ultraviolet regulator as far as the field theory is concerned.

Since (\ref{RegulatedKG}) is regulated, our next step must be to renormalize by adding suitable counterterms. This is called ``holographic renormalization" \cite{Bianchi:2001de, Bianchi:2001kw} (for a review see \cite{Skenderis:2002wp}). We will demonstrate how it works in a particular example: let $d=3$ so we are dealing with $AdS_4/{\rm CFT}_3$ and take a scalar field with mass-squared $m^2 L^2 = -2$. (This example is relevant for the gravity dual to ABJM theory.) Now $\Delta_+ = 2$, $\Delta_- = 1$ and the asymptotic behavior of the scalar is
\begin{eqnarray}
\phi(r \to \infty, \vec{x}, t) = {\alpha(\vec{x},t)L^2 \over r} + {\beta(\vec{x},t) L^4\over r^2} + \cdots \,.
\end{eqnarray}
Plugging this into the KG action, the bulk term vanishes and we are left with the boundary term
\begin{eqnarray}
S_{\rm KG}  = {L^2 \over 4 \kappa^2}\int d^3x \left(   {R \alpha^2 \over L^2} + 3   \alpha \beta
\right) \,,
\end{eqnarray}
which is implicitly evaluated at $r=R$, diverging as $R \to \infty$. We will deal with this divergence by adding a new piece to the action, a boundary counterterm,
\begin{eqnarray}
\label{BdyCounter}
S_{\rm bdy} = - {1 \over 4\kappa^2} \int d^3x \sqrt{-h}\,  \phi^2 \,.
\end{eqnarray}
The divergence of this boundary term cancels the divergence of the bulk action, leaving us with the finite result
\begin{eqnarray}
S_{\rm KG} + S_{\rm bdy} = {L^2 \over 4 \kappa^2} \int d^3x \,  \alpha \beta \,.
\end{eqnarray}
Our choice of boundary counterterm is also intimately related to our boundary conditions (\ref{RegularQuant}) constraining $\alpha(\vec{x}, t)$ (for a nice discussion of this, see \cite{Marolf:2006nd}). The relationship comes from making sure our solution is a true extremum of the action, boundary terms included. Consider varying the field
\begin{eqnarray}
\phi(r, \vec{x}, t) \to \phi(r, \vec{x}, t) + \delta \phi(r, \vec{x}, t) \,.
\end{eqnarray}
This induces variations $\delta \alpha(\vec{x}, t)$ and $\delta \beta (\vec{x}, t)$. We want the action, evaluated on a solution, to be stationary under such a variation. 
The variation is
\begin{eqnarray}\nonumber
\delta S_{\rm KG} + \delta S_{\rm bdy} &=&  {1 \over 2 \kappa^2}\int d^4x \sqrt{-g} \, \delta \phi \left( \square \phi +2 \phi \right) \\ &+& {L^2 \over 2 \kappa^2}\int d^3x \left( {R\over L^2} (1 - 1) \alpha \delta \alpha  + (1 -1) \alpha \delta \beta
+ (2 - 1) \beta \delta \alpha \right) \,, 
\end{eqnarray}
Solving the Klein-Gordon equation ensures the bulk part is zero. Moreover, we can see that the boundary term (\ref{BdyCounter}) cancels a divergent part of the variation, as well as a finite $\delta \beta$ part. We are left with
\begin{eqnarray}
\label{SVariation}
\delta S_{\rm KG} + \delta S_{\rm bdy} = {L^2 \over 2 \kappa^2 }\int d^3x \,
 \beta \delta \alpha  \,.
\end{eqnarray}
For a general boundary condition this would not vanish, but for our boundary condition (\ref{RegularQuant}),
$\alpha$ is not allowed to fluctuate and so its variation must be zero. Thus indeed the solution is stationary under the full, bulk plus boundary action.

Thus the boundary counterterm has done two things for us:
\begin{enumerate}
	{\item Made the total action finite as the regulator is removed, and}
	 {\item Made the total action stationary for solutions once our boundary conditions are imposed.}
\end{enumerate}
The variation of the action (\ref{SVariation}) also provides our path to correlation functions. Let's say we want to calculate the one-point function of the field theory operator ${\cal O}$ dual to $\phi$. To get this we vary with respect to the source:
\begin{eqnarray}
\langle {\cal O}(\vec{x}, t) \rangle = {1 \over i} {\delta \over \delta J(\vec{x}, t)}  \left\langle \exp{i \int d^dx \, J(\vec{x}, t) {\cal O}(\vec{x}, t)} \right\rangle  \Bigg|_{J = 0}\,,
\end{eqnarray}
where we assume we have normalized the partition function to 1 in the absence of sources.
But using the correspondence, this can be translated into a statement about the gravity action and its response to varying its boundary conditions: 
\begin{eqnarray}
\langle {\cal O}(\vec{x}, t) \rangle = {1 \over i} {\delta \over \delta \alpha(\vec{x}, t)}  e^{i S_{\rm grav}} \Bigg|_{\alpha = 0} =  {\delta  S_{\rm grav}\over \delta \alpha(\vec{x}, t)} \Bigg|_{\alpha = 0}\,.
\end{eqnarray}
In our example of a single scalar field, (\ref{SVariation}) tells us this is simply
\begin{eqnarray}
\label{OnePointReg}
\langle {\cal O}(\vec{x}, t) \rangle = {L^2 \over 2 \kappa^2 }\beta(\vec{x}, t) \,.
\end{eqnarray}
Thus for our example, we have 
\begin{eqnarray}
\phi(r \to \infty, \vec{x}, t) = {L^2 J(\vec{x},t) \over r} + { 2 \kappa^2 L^2 \langle{\cal O} (\vec{x},t) \rangle\over r^2} + \cdots \,.
\end{eqnarray}
Just as the leading term near the boundary corresponds to the source for the dual operator, and is constrained by our boundary conditions, the subleading term which is allowed to fluctuate corresponds to the expectation value of the dual operator. This kind of relationship generalizes to all fields in all dimensions. We may think of the two terms as the stimulus and the response: the source term $\alpha(\vec{x}, t)$ pokes the system, and combined with the initial conditions this determines how the system responds $\beta(\vec{x}, t)$. We will discuss how the source communicates its influence to the response in section~\ref{CorrelationSec}, where we discuss higher-point correlation functions.

\subsection{Alternate quantization and the other relevant operators} 

By allowing our scalar mass-squared to go down to the Breitenlohner-Freedman bound, we have been able to find gravity duals for operators with dimension down to $\Delta = d$. However, some physical operators in known systems have dimension smaller than this. How can we realize these operators in the gravity dual?

The answer turns out to be to change our boundary conditions. For a restricted range of $m^2$ values down to (but not including) the BF bound,
\begin{eqnarray}
\label{AltRange}
- {d^2 \over 4} < m^2 L^2 \leq - {d^2 \over 4} + 1 \,,
\end{eqnarray}
it is possible to exchange the roles of $\alpha$ and $\beta$: now we will take $\beta$ to be fixed, and allow $\alpha$ to fluctuate.

Since we are changing our boundary conditions, we will have to change our boundary terms, as well.   Our example of the field with $d=3$ and $m^2 L^2 = -2$ lies in the range (\ref{AltRange}). Consider the alternate boundary terms:
\begin{eqnarray}
S_{\rm bdy, alt} = {1 \over 4 \kappa^2} \int d^3x \sqrt{-h} \, \phi^2 + {1 \over 2 \kappa^2}\int d^3x \sqrt{-h} \, \phi n^\mu \partial_\mu \phi \,.
\end{eqnarray}
We have changed the sign of the original boundary term, and added a second term. This combination also renders the total action finite:
\begin{eqnarray}\nonumber
S_{\rm KG} + S_{\rm bdy, alt} &=& {L^2 \over 4 \kappa^2}\int d^3x \left( {R\over L^2} \left( 1 + 1-2 \right) \alpha^2  + \left( 3 +2 - 6 \right) \alpha \beta  \right)  \\ &=& - {1 \over 4 \kappa^2 L^4} \int d^3x \,\alpha \beta\,,
\end{eqnarray}
and leads to the variation of the action
\begin{eqnarray}
\label{AltSVariation}
\nonumber
\delta S_{\rm KG} + \delta S_{\rm bdy, alt} &=& {L^2 \over 2 \kappa^2 } \int d^3x \left( {R \over L^2}(1 +1 - 2) \alpha \delta \alpha + (1 + 1 - 3) \alpha \delta \beta
+ (2 +1 - 3) \beta \delta \alpha \right) \\
&=& - {L^2 \over 2 \kappa^2 }\int d^3x  \, \alpha \delta \beta \,.
\end{eqnarray}
Thus the solutions leave the action stationary when we constrain $\beta(\vec{x}, t)$,
\begin{eqnarray}
\label{AltSource}
\beta(\vec{x}, t) =J_{\rm alt}(\vec{x}, t) \,,
\end{eqnarray}
and allow $\alpha(\vec{x}, t)$ to fluctuate.\footnote{When working in asymptotically AdS space, the relation (\ref{AltSource}) can become contaminated by $\alpha$ terms unless one works in ``Fefferman-Graham" coordinates where $g_{rr} = L^2/r^2$ exactly.} The alternate quantization variation (\ref{AltSVariation}) then implies
\begin{eqnarray}
\langle {\cal O}_{\rm alt}(\vec{x}, t) \rangle =- {L^2 \over 2 \kappa^2}\alpha(\vec{x}, t) \,.
\end{eqnarray}
Now $\alpha$ and $\beta$ have switched roles. Thus the dual operator ${\cal O}_{\rm alt}$ must have the scaling dimension of $\alpha$, which means $\Delta_{{\cal O}_{\rm alt}} =  \Delta_-$.
\begin{figure}
\begin{center}
\includegraphics[scale=0.7]{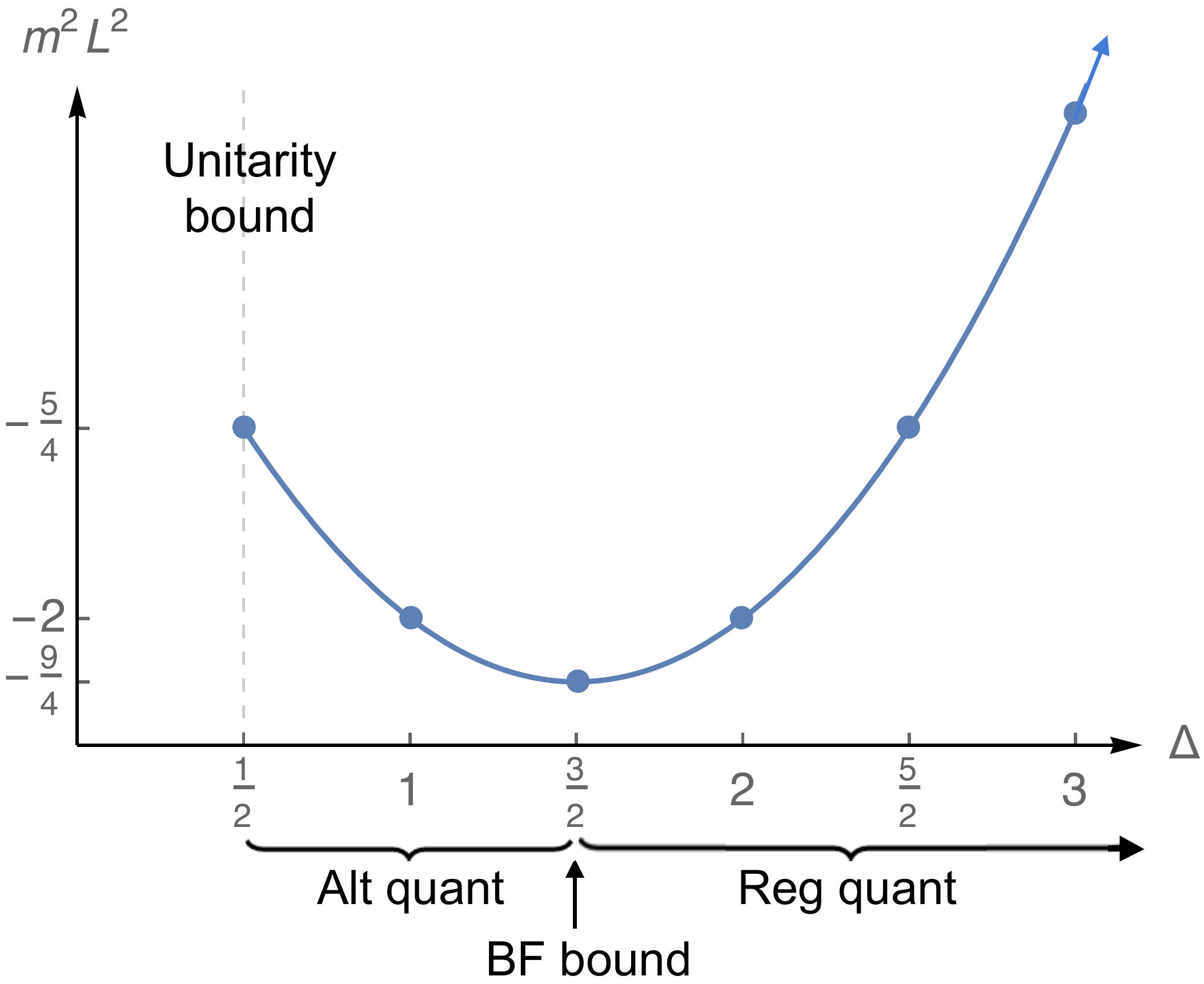}
\caption{The plot of mass-squared $m^2$ of the gravity scalar $\phi$ versus the dimension $\Delta$ of the dual field theory operator ${\cal O}$, for AdS$_4$/CFT$_3$ ($d=3$). Heavy dots indicate cases with integer or half-integer dimension. Values of $\Delta$ associated to the regular quantization and alternate quantization are indicated, as is the value of $\Delta$ that saturates the BF bound $m^2 L^2 = -9/4$, and the unitarity bound $\Delta \geq 1/2$. ${\cal N}=8$ gauged supergravity in four dimensions has 70 scalars at $m^2L^2=-2$, 35 in alternate quantization dual to the $\Delta = 1$ operator Tr $X^2$, and 35 in regular quantization dual to the $\Delta = 2$ operator Tr $\lambda^2$.
\label{fig:DimensionMass}}
\end{center}
\end{figure}

Using masses over the range (\ref{AltRange}), this brings us all the way down to $\Delta = d/2-1$. As it turns out, this is as low as you can go: unitarity and the conformal algebra of a CFT implies $\Delta \geq d/2-1$. The gravity side also stops there: if the mass-squared is outside the range (\ref{AltRange}), there are no boundary terms compatible with the alternate boundary conditions to render the action finite, and we must use the regular quantization.

The alternate quantization may seem a little esoteric, but it is physically essential. Consider the example of M-theory on $AdS_4 \times S^7$, dual to ABJM theory. In the gravity limit M-theory becomes eleven-dimensional supergravity, and we can reduce this theory on $S^7$, producing towers of fields. The ``bottom" set of fields in these towers, consisting of the four-dimensional graviton and its superpartners, constitutes the fields of four-dimensional, ${\cal N}=8$ gauged supergravity. This theory has 70 scalars, all with $m^2 L^2 = -2$. It turns out that supersymmetry requires half of these scalars to use the regular quantization, and the other half the alternate quantization. This fits perfectly with the dual ABJM theory, where there are 35 scalar bilinears Tr $X^2$ with $\Delta = 1$, and 35 fermion bilinears Tr $\lambda^2$ with $\Delta = 2$. We see the importance of the quantization of the boundary conditions: even if we kept the same fields on the gravity side, changing the boundary conditions would change the operator content of the field theory dual. The same bulk gravity action with different boundary conditions, and hence different boundary counterterms, is truly a different theory.

To illustrate these ideas, in figure~\ref{fig:DimensionMass} we plot the relationship between the conformal dimension $\Delta$ and the mass-squared $m^2L^2$ for a scalar with $d=3$. We indicate the range of $\Delta$ corresponding to regular quantization, the range of $\Delta$ corresponding to alternate quantization, and the value $\Delta = 3/2$ where the associated mass saturates the BF bound $m^2L^2 = -9/4$.

\subsection{Fields with spin}

Bosonic fields on the gravity side like the metric and gauge fields also satisfy second order equations, and analogously to the scalar described above, one independent solution near the boundary corresponds to a source for the dual operator, while the other solution is proportional to the one-point function.

As mentioned before when talking about symmetries, vector fields on the gravity side are dual to conserved currents on the field theory side:
\begin{eqnarray}
A_\mu \longleftrightarrow J^\mu \,.
\end{eqnarray}
Let us outline the holographic renormalization of the $A_\mu$ field, briefly since many aspects are analogous to the case of the scalar. The quadratic Maxwell action
\begin{eqnarray}
S_{\rm Max} = {1 \over 2 \kappa^2} \int d^{d+1}x \sqrt{-g} \, \left( - {1 \over 4} F_{\mu\nu}F^{\mu\nu}\right)\,,	
\end{eqnarray}
with $F_{\mu\nu} \equiv \partial_\mu A_\nu - \partial_\nu A_\mu$ leads to the equation of motion
\begin{eqnarray}
{1 \over \sqrt{-g}} \partial_\mu \sqrt{-g} g^{\mu\alpha} g^{\nu\beta} F_{\alpha \beta} = 0 \,.	
\end{eqnarray}
In the gauge $A_r = 0$, the solution to the equations of motion near the boundary is
\begin{eqnarray}
\label{VecFieldExpand}
A_i(r, \vec{x},t ) = \alpha_i(\vec{x},t ) L  + {\beta_i(\vec{x},t ) L^{2d-3} \over r^{d-2}} + \ldots \,,
\end{eqnarray}
where $\partial^i \beta_j=0$, and $f_{ij} \equiv \partial_i \alpha_j - \partial_j \alpha_i$ and $g_{ij} \equiv \partial_i \beta_j - \partial_j \beta_i$ are constrained by $\partial^i f_{ij} = \partial^i g_{ij} = 0$. To see the scaling of $\alpha_i$ and $\beta_i$, pass to locally flat coordinates:
\begin{eqnarray}
A_{\hat{\imath}}(r, \vec{x},t ) = {\alpha_i(x) L^2 \over r}  + {\beta_i(x) L^{2d-2} \over r^{d-1}} + \ldots \,,
\end{eqnarray}
where we see that $\alpha_i(\vec{x},t )$ should be dual to a source of dimension 1, and $\beta_i(\vec{x},t )$ to an operator of dimension $d-1$; again we have inserted factors of $L$ so the scaling dimensions match the engineering dimensions.

The action turns out to be finite without the addition of any boundary term, and reduces on the equation of motion to 
\begin{eqnarray}
S_{\rm Max} = (d-2){L^{d-1} \over 4 \kappa^2} \int d^dx \, \alpha^i \beta_i \,,
\end{eqnarray}
while the variation of the action becomes
\begin{eqnarray}
\delta S_{\rm Max} = (d-2){L^{d-1} \over 2 \kappa^2} \int d^dx \, \delta \alpha^i \beta_i \,.
\end{eqnarray}
Thus we identify $\alpha_i(\vec{x},t)$ as the fixed source for the current, and the one-point function is
\begin{eqnarray}
\langle J^i(\vec{x},t) \rangle = (d-2) {L^{d-1} \over 2 \kappa^2} \eta^{ij} \beta_j(\vec{x},t) \,,
\end{eqnarray}
completing our discussion of the spin-1 field.

Finally, the (unique) metric is dual to the (also unique) energy-momentum tensor:
\begin{eqnarray}
g_{\mu\nu} \longleftrightarrow T^{\mu\nu} \,.
\end{eqnarray}
We will not go through the holographic renormalization of the metric tensor, which is somewhat complicated and involves adding boundary terms with a geometric meaning. For a further discussion of this, including the holographic Weyl anomaly, see \cite{Skenderis:2000in}.

Thus we have described how bosonic fields match up with their corresponding operators. We will say a little about fermionic fields in the final section.

\section{String theory origins of AdS/CFT}

Before diving into studying correlation functions, let's take a side trip to think about how string theory motivated the AdS/CFT correspondence, and how this relates to the two distinct approaches to applying gravity systems to learn about gauge theories, which one can call ``top-down" and ``bottom-up". Our discussion of string theory will just skim the surface, but hopefully provide enough of a flavor to communicate the the role it plays. To explore this more deeply, see for example \cite{Aharony:1999ti, Klebanov:2000me, DHoker:2002nbb}.

\subsection{D-branes, M-branes and the gauge/gravity correspondence}

String theory is the result of quantizing one-dimensional relativistic objects, called strings. Strings come in two varieties: closed strings, which loop back on themselves, and open strings, which have endpoints. The quantization of closed strings naturally leads to the dynamics of gravity (plus other fields), while the quantization of open strings leads to the dynamics of gauge theory (again, plus other fields). String theory needs to live in an unexpectedly large number of dimensions in order to avoid quantum inconsistencies: for the supersymmetric version of the theory, which contains fermionic excitations, this number is ten.

A fundamental property of string theory is that a little piece of string does not ``know" whether it is part of a closed string or an open string unless it is at an endpoint; thus, the difference between gravity and gauge theory comes solely from the boundary conditions, as closed strings can have both left- and right-moving oscillations, while in open strings half the degrees of freedom are projected out and we are left with standing waves. In some sense, string theory is telling us that gravity is like gauge theory squared \cite{Kawai:1985xq}, and the two kinds of physics are intimately connected.

The original understanding of open strings imposed Neumann boundary conditions on their endpoints in all directions, resulting in strings that could end anywhere. After some time, it was realized that one can instead impose Dirichlet boundary conditions in some directions, restricting the endpoints to live on certain lower-dimensional surfaces in spacetime \cite{Dai:1989ua}. The next step was realizing that these surfaces that open strings can end on, so-called ``Dirichlet-branes" or ``D-branes", behave themselves as dynamical membrane objects that can carry energy, move around and interact, and have a particular kind of ``Ramond-Ramond" charge  \cite{Polchinski:1995mt}. At low energies, the dynamics of the open strings living on the branes are a lower-dimensional gauge theory, and the fact that these strings can interact with closed strings moving throughout space indicates that the branes gravitate, as well as being sources for higher-index generalizations of abelian gauge fields.

The birthplace of AdS/CFT was in these D-branes. Consider a stack of $N$ D3-branes\footnote{A $Dp$-brane traditionally denotes a brane with $p$ spatial dimensions, and thus $(p+1)$ spacetime dimensions.} on top of each other; these branes exist within the string theory called type IIB. The ability of each open string endpoint to end on any of the $N$ branes gives rise to the $N^2$ degrees of freedom of a non-abelian gauge theory, in this case the maximally supersymmetric gauge theory in four dimensions, ${\cal N}=4$ Super-Yang-Mills theory with $SU(N)$ gauge group. On the other hand, the energy in the branes curves space around them, altering the geometry. The breakthrough of Maldacena \cite{Maldacena:1997re} was to propose that the gauge theory living {\em on} the branes was {\em exactly equivalent} to the gravitational dynamics in the geometry {\em very close} to the branes; the two descriptions are redundant with each other. The near-horizon geometry of a stack of D3-branes is $AdS_5 \times S^5$, and from this emerged the proposal that type IIB string theory on this spacetime was exactly dual to ${\cal N}=4$ Super-Yang-Mills.

From this origin of the correspondence, details of the field-operator map emerge naturally. In the original brane picture, there are couplings between the open string modes living on the brane, and closed string excitations propagating through spacetime. These couplings imply a connection between gauge-invariant field theory operators and gravity fields. Thus in this example, the AdS/CFT dictionary can be derived.

While there are D-branes of other dimensionalities, the D3-branes are special in hosting an exactly conformal field theory. It turns out there are two other branes that have this property, though they are not exactly D-branes. One other remarkable idea that along with D-branes propelled the so-called ``second superstring revolution" of the mid-to-late 1990's was Witten's proposal that the strongly coupled limit of (ten-dimensional) type IIA string theory was not a string theory at all, but an eleven-dimensional theory containing membrane degrees of freedom, dubbed M-theory, whose low-energy limit was the already-discovered eleven-dimensional supergravity \cite{Witten:1995ex}. M-theory is now understood (albeit imperfectly) as being part of the web of dualities relating various string theories. It has no strings, but does have branes of its own, M2-branes and M5-branes. The low-energy field theories living on these branes are the three-dimensional ABJM theory, and the six-dimensional (2,0) theory; along with ${\cal N}=4$ SYM, these theories form the fundamental trio of maximally supersymmetric, exactly conformal field theories in more than two dimensions. The field-operator map could be read off for these theories as well--- although the (2,0) theory, in particular, is still not as well understood from the field theory point of view.

These theories and their generalizations are called {\em top-down} realizations of AdS/CFT. Because they are motivated from string/M-theory, their field operator dicitonaries are known, and the footing they stand on is relatively firm.

\subsection{Top-down vs bottom-up}

We now come to the two different approaches for studying field theories using gravity: ``top-down" and ``bottom-up".

In the top-down case, one uses a known gravity theory/field theory pair coming from string/M-theory. This has several advantages:
\begin{itemize}
{\item Because the dictionary is known precisely, you know exactly which operators in which field theory you are talking about.}
{\item Because the dual theories come from string theory, you can be confident --- or at least, {\rm more} confident --- that there aren't hidden pathologies affecting the system. }	
\end{itemize}
But there are also drawbacks:
\begin{itemize}
	{\item You are limited to studying one of the field theories that is known to have a string theory dual. These usually have a lot of supersymmetry and may not look exactly like known systems realized in nature.}
	{\item These systems are often complicated. For example, type IIB supergravity on $AdS_5 \times S^5$ has an infinite set of fields; even truncating to the graviton and its superpartners, one is left with five-dimensional ${\cal N}=8$ gauged supergravity, a theory with 15 gauge fields and 42 scalars, among other bose and fermi fields.}
\end{itemize}
On the flip side, one can pursue a ``bottom-up" model: here one just writes down a simple gravity theory with whatever properties are desired. Now instead of using a complex set of fields pre-supplied by string theory, you just use what you want. Put in a graviton. Need a conserved current? There's a $U(1)$ gauge field. Interested in a charged condensate? Add a charged scalar. The interactions can be tweaked on demand.  

Now the advantages have flipped:
\begin{itemize}
	{\item The model is flexible, and can contain whatever fields and dynamics you want.}
	{\item The model can be simple, with no extraneous fields or complicated couplings.}
\end{itemize}
But so have the disadvantages:
\begin{itemize}
	{\item Since string theory didn't give it to you, you don't know what the dual field theory is. At best you can say it's some large N gauge theory at strong coupling, with particular symmetries and an operator spectrum you created. You don't have a Lagrangian or even a list of fundamental fields.}
	{\item There might be some hidden pathology or issue with the theory that you can't see.}
\end{itemize}
In practice, both of these approaches are valuable. The bottom-up approach can focus in directly on a particular desired property, and tune interactions and couplings to get exactly the phenomenon that one wants. On the other hand, the top-down approach makes precise predictions about known field theories and is on a firmer footing. Moreover, sometimes certain kinds of interactions or dynamics that you might not have thought of by yourself can be offered to you by the top-down model.

In our examples in the second half of the lectures, we will explore both top-down and bottom-up approaches, and hopefully see the value of both.

\section{Correlation Functions and RG flow geometries}
\label{CorrelationSec}

We will now turn to the study of two-point correlation functions. We will discuss what these look like in an exactly conformal field theory dual to AdS space. Then, we will study a few new geometries that are only asymptotically AdS, discuss how the variation of their fields over the radial coordinate is associated to the breaking of scale invariance, and then discuss two-point functions in these systems as well.

\subsection{Two-point functions and boundary conditions}

Higher-point correlation functions can also be calculated from the gravity side. Consider a two-point function; this can be calculated as the variation with respect to the source of the one-point function, with sources not turned off until the end,
\begin{eqnarray}
\langle {\cal O}(x) {\cal O}(y) \rangle  = {1 \over i} {\delta \langle {\cal O}(x)\rangle_J \over \delta J(y)} \Bigg|_{J=0} \,,
\end{eqnarray}
where $x =\{\vec{x},t\}$.
For a scalar in the regular quantization, the generalization of the relations (\ref{RegularQuant}) and (\ref{OnePointReg}) to arbitrary $d$ and $\Delta$ are 
\begin{eqnarray}
\label{GeneralRegQuant}
J(x) = \alpha(x)   \,, \quad \quad   
\langle{\cal O}(x)\rangle =  (2 \Delta - d) {L^{d-1} \over 2 \kappa^2}  \beta(x) \,,
\end{eqnarray}
which imply the two-point function\footnote{The alternate quantization version of (\ref{GeneralRegQuant}) switches $\alpha$ and $\beta$ and adds a sign to the expression for $\langle {\cal O} \rangle$.}
\begin{eqnarray}
i\langle {\cal O}(x) {\cal O}(y) \rangle = (2 \Delta-d) {L^{d-1} \over 2 \kappa^2} {\delta \beta(x) \over \delta \alpha(y)} \Bigg|_{\alpha = 0}\,.
\end{eqnarray}
In the near-boundary expansion, $\alpha(x)$ and $\beta(x)$ are independent. What causes one to depend on the other? To relate them, one must solve for $\phi(x,r)$ throughout the bulk, in general imposing a boundary condition far from the boundary, at the deep interior (infrared end) of the geometry. One will then have a functional relation like
\begin{eqnarray}
\label{Kernel}
\beta(y) = \int d^dx K(x-y) \alpha(x) + {\cal O}(\alpha^2) \,,
\end{eqnarray}
with a kernel $K(x-y)$ proportional to the two-point function. Since translation invariance guarantees the two-point function depends only on the difference $x-y$, we can detangle the integral relation (\ref{Kernel}) by passing to momentum space, where we find the simple expression
\begin{eqnarray}
i\langle {\cal O}(p) {\cal O}(-p)\rangle =  {1 \over (2\pi)^d}  {\langle{\cal O}(p)\rangle \over J(p)}= {2 \Delta-d \over (2\pi)^d}  {L^{d-1} \over 2 \kappa^2} {\beta(p) \over \alpha(p)}\,,
\end{eqnarray}
neglecting higher order terms in (\ref{Kernel}).
We poke the gravity geometry with a source by imposing the $\alpha$ boundary condition, and the field responds with $\beta$; the two point-function is just the ratio.

Thus interaction terms are in general not necessary to study two-point functions, since we only need the linearized equations to determine (\ref{Kernel}) to leading order. For higher-point functions we must go beyond linearized order in the equations of motion, and interactions start to play a role.

Having established the structure of the two-point function, the task becomes to solve the scalar equation of motion in the appropriate background, imposing a boundary condition in the deep interior, and then read off the results at the boundary. The boundary condition in the interior thus becomes very important; which condition shall we pick?

For zero temperature backgrounds, the linearized equation can be solved continuing to Euclidean space. In the deep interior there is generally one diverging solution and one regular solution, and the prescription is to choose the regular solution. We say we choose {\em regular boundary conditions}. We shall see some examples of this in the rest of the section. At finite temperature solutions can have different behavior in the deep interior, and then we need to find  a different boundary condition, as we will discuss in section~\ref{RealTimeSec}.

\subsection{$N$-scaling of correlation functions}
\label{NScalingSec}

The one-point functions all contain a factor $L^{d-1}/\kappa^2$ coming from the overall normalization of the gravity action. This ratio of two gravitational quantities must reduce to something purely field-theoretic. In the top-down models where we understand the dictionary directly, we can evaluate this explicitly, and discover it reduces to the scaling of the correlator with a power of $N$.

In the case of type IIB on $AdS_5 \times S^5$ dual to ${\cal N}=4$ Super -Yang-Mills, one has
\begin{eqnarray}
L^4 = 4 \pi g_s N {\alpha'}^2 \,,\quad\quad\quad{1 \over \kappa^2}  = {L^5 \over 64 \pi^4 g_s^2 {\alpha'}^4} \,.
\end{eqnarray}
We see that the string theory parameters $g_s$ (the string coupling) and $\alpha'$ (the string length squared) cancel out of our ratio,
\begin{eqnarray}
\label{LKappa}
{L^3 \over \kappa^2} = {N^2 \over 4 \pi^2} \,.
\end{eqnarray}
Thus correlation functions in ${\cal N}=4$ SYM go like $N^2$ in the large-$N$ limit; this is a known result in large-$N$ gauge theories.
In a bottom-up construction, one cannot calculate these factors from first principles. Nonetheless, in a bottom-up AdS$_5$ model, one generally imagines that the dual field theory is some non-Abelian gauge theory, with the implication that $L^3/\kappa^2 \propto N^2$ still holds.

In more general top-down cases, if the gravity theory is of the form $AdS_{d+1} \times S^q$ associated to the backreaction of a set of $N$ $d$-dimensional branes, the AdS radius and gravitational constant will take the form
\begin{eqnarray}
L^{q-1} \propto N \ell_P^{q-1} \,, \quad \quad \quad {1 \over \kappa^2} \propto {L^q\over \ell_P^{d+q-1} }\,,
\end{eqnarray}
where $\ell_P$ is the Planck length of the higher-dimensional theory, defined by the higher-dimensional gravitational constant $\kappa_{\rm higher}^2 \propto \ell_P^{d+q-1}$. For ABJM theory and the six-dimensional (2,0) theory, one finds
\begin{eqnarray}
{L^2 \over \kappa^2} \sim N^{3/2} \quad \hbox{(ABJM)}\,, \quad \quad \quad{L^5 \over \kappa^2} \sim N^3 \quad \hbox{(2,0)} \,,
\end{eqnarray}
which are the characteristic powers of $N$ associated with these theories.

\subsection{Two-point functions in AdS space}

In pure AdS space, the massive Klein-Gordon equation is most easily solved in terms of the $z \equiv L^2/r$ coordinates. Working in Euclidean space and Fourier transforming the $d$ boundary coordinates $x$ to a momentum $p$, it takes the form
\begin{eqnarray}
z^2 \phi'' - (d-1) z \phi' - z^2 p^2 \phi - m^2 L^2 \phi = 0\,.
\end{eqnarray}
This has solution in terms of modified Bessel functions,
\begin{eqnarray}
\phi(z, p) =  c_1  z^{d/2}K_\nu ( pz) + c_2 z^{d/2} I_\nu (p z) \,, \quad \quad \nu \equiv \sqrt{{d^2 \over 4} + m^2 L^2 }= \Delta_+ - 
{d\over2} \,.
\end{eqnarray}
In the deep interior $z \to \infty$, $I_\nu$ diverges as $e^{z}$ while $K_\nu$ is regular, going like $e^{-z}$. We satisfy our regularity prescription  for the boundary condition by keeping the $K_\nu$ solution,
\begin{eqnarray}
\phi(r, p) = r^{-d/2} \, K_\nu \Big( {p L^2 \over r }\Big)  \,.
\end{eqnarray}
Expanding this near the boundary, we find the ratio of the $\alpha$ and $\beta$ terms give for integer $\nu$ the 2-point function,
\begin{eqnarray}
\label{TwoPointLog}
\langle {\cal O}(p) {\cal O}(-p)\rangle \sim (p^2)^\nu \log p^2 \,,\quad \quad \hbox{integer}\ \nu \,, 
\end{eqnarray}
and for non-integer $\nu$,
\begin{eqnarray}
\label{TwoPointNoLog}
\langle {\cal O}(p) {\cal O}(-p)\rangle \sim (p^2)^\nu \,, \quad \quad \hbox{non-integer}\ \nu \,,
\end{eqnarray}
In either case, the Fourier transform gives us in position space\footnote{In general (\ref{TwoPointLog}) will have additional analytic $p^{2\nu}$ terms, which become scheme-dependent contact terms in position space.}
\begin{eqnarray}
\langle {\cal O}(x) {\cal O}(x') \rangle = {C \over |x-x'|^{2\Delta}} \,,
\end{eqnarray}
up to a constant $C$ related to the normalization of ${\cal O}$.
This is exactly the functional form required by conformal invariance in a CFT.

Whether from the log or the non-integer power $\nu$, the correlation functions (\ref{TwoPointLog}), (\ref{TwoPointNoLog}) have a non-analyticity in $p^2$, which we can view as creating a branch cut at the origin in the complex $p^2$ plane. This has the interpretation of a continuum of states all the way down to zero energy; since this is an exactly conformal field theory with no preferred scale,  excitations of all energies are possible. 

Three and higher point functions can also be computed, and have the forms conformal field theory demands. We will not continue down the road of higher point functions, but there is a lovely story there; see for example \cite{DHoker:2002nbb}.

\subsection{RG flow geometries}

Let's think about some slightly more complicated geometries than pure AdS: consider in the background a single scalar field varying in the radial direction, the ``active" scalar,
\begin{eqnarray}
\label{RGFlowScalar}
\phi = \phi(r) \,.
\end{eqnarray}
In general this leads to an asymptotically AdS metric of the form (\ref{AsymptoticAdS}) with $h(r) = 1$:
\begin{eqnarray}
\label{RGFlow}
ds^2 = e^{2A(r)} \eta_{ij} dx^i dx^j + e^{2B(r)} dr^2 \,.
\end{eqnarray}
These geometries preserve $d$-dimensional Poincar\'e invariance, but break scale invariance. Thus the state of the dual field theory evolves as one runs from the ultraviolet to the infrared, and spacetimes of the form (\ref{RGFlowScalar}), (\ref{RGFlow}) are known as ``renormalization group flow" (RG flow) geometries. As we will see, depending on the behavior of $\phi(r \to \infty)$ such a geometry sometimes corresponds to a non-vacuum state of a CFT (spontaneous breaking of conformal invariance), and sometimes it corresponds to a new, non-CFT theory altogether (explicit breaking of conformal invariance). 

We will illustrate RG flow geometries in the particular context of the five-dimensional ${\cal N}=8$ gauged supergravity theory, which is a truncation of type IIB supergravity on $AdS_5 \times S^5$ and hence is dual to ${\cal N}=4$ Super-Yang-Mills (specifically, to its lowest dimension operators that remain at strong coupling). The bosonic modes include the metric, the $SO(6)$ 15 gauge fields from the Kaluza-Klein reduction on the $S^5$, and 42 scalars. We summarize these scalars, their 10-dimensional origin, their $SO(6)$ quantum numbers, mass $m^2$, and associated dual operator and dimension $\Delta$ in the table,

\begin{center}
\begin{tabular}{|c|c|c|c|c|}
	\hline 10D SUGRA fields & $SO(6)$ & $m^2 L^2$ & $\Delta$ & ${\cal N}=4$ operator \\ \hline
	$g_{\mu\nu} + F_5$ & ${\bf 20'}$ & $-4$ & 2 & Tr $X^{(i} X^{j)}$\\
	$H_3$, $F_3$ & $ {\bf 10} \oplus {\bf \overline{10}}$ & $-3$ & 3 & Tr $\lambda \lambda$ \\
	IIB dilaton & ${\bf 1} \oplus {\bf 1}$ & $0$ & $4$ & ${\cal L} = {\rm Tr}\ (F_{\mu\nu}F^{\mu\nu} + \cdots)$ \\ \hline
\end{tabular}
\end{center}
Here $F_5$, $F_3$ and $H_3$ are antisymmetric tensor field strengths of type IIB supergravity, and $X$, $\lambda$ and $F$ are scalars, fermions and  field strengths of ${\cal N}=4$ SYM, transforming in the ${\bf 6}$, ${\bf 4}$ and ${\bf 1}$ of $SO(6)$, respectively; the ${\bf 20'}$ representation of $SO(6)$ has the prime because there is already a representation called ${\bf 20}$. For more details on the fields and operators and their relations, see \cite{DHoker:2002nbb}.

\bigskip
\noindent
\underline{\bf Coulomb branch flow}

\noindent
Consider a particular RG flow geometry with an ``active" scalar $\phi_{\bf 20'}(r)$ in the ${\bf 20'}$ of $SO(6)$, which has dual operator Tr $X^2$ with $\Delta = 2$. There is a supersymmetric solution preserving half the total supersymmetry (16 supercharges) with the form \cite{Freedman:1999gk}
\begin{eqnarray}
e^{A(r)} = {r \over L }\left( 1 + {\ell^2\over r^2} \right)^{1/6}\,, \quad
e^{B(r)} = {L \over r}\left( 1 + {\ell^2\over r^2} \right)^{-1/3}\,,  \quad
\phi_{\bf 20'}(r) =  -\sqrt{2\over 3} \log \left( 1 + {\ell^2\over r^2} \right) \,,
\end{eqnarray}
in terms of a parameter $\ell$, which when set to zero returns us to AdS space. This solution breaks $SO(6) \to SO(4) \times SO(2)$. The near-boundary expansion for the scalar is
\begin{eqnarray}
\phi_{\bf 20'} = 0 \times {\log r \over r} + {\ell^2 \over r^2} + \ldots \,.
\end{eqnarray}
Here we have explicitly written how the coefficient of the ``source" term in the near-boundary scalar expansion is zero (note this is a BF-bound saturating scalar with the form (\ref{BFBoundSaturating})). Hence there is zero source turned on, but there is an expectation value
\begin{eqnarray}
\langle {\rm Tr}\ X^2\rangle \sim \ell^2\,.
\end{eqnarray}
The interpretation for this geometry is that it is a state in ${\cal N}=4$ SYM where some expectation values have been turned on for the scalars $X$, spontaneously breaking conformal symmetry and moving out onto the Coulomb branch.\footnote{ ``Coulomb branch" is a term in the lingo of supersymmetric gauge theories indicating a scalar in the adjoint representation of the gauge group, a superpartner of the gauge fields, has gotten an expectation value.}

This geometry has a singularity in the IR, and one might be concerned that this singularity renders the geometry problematic. However, it is known that this spacetime is the five-dimensional reduction of a well-behaved ten-dimensional spacetime with a disc of D3-brane sources. Consequently, the spacetime is considered an acceptable one despite its singularity.

Let us take a look at a two-point function in this geometry. Because the active scalar $\phi_{\bf 20'}$ is nonzero in the background, its fluctuations mix with fluctuations of the metric trace, and are more complicated to deal with. A simpler thing to look at is the fluctuations of the physical (transverse traceless) modes of the graviton; in any RG flow spacetime these always satisfy the massless Klein-Gordon equation.

The solution for the linearized Klein-Gordon equation for a massless scalar $h$ can be found in terms of hypergeometric functions \cite{Freedman:1999gk}, 
\begin{eqnarray}
h  = \left( r^2 \over r^2 + \ell^2 \right)^a \: {}_2F_1 \Big( a,a;2+2a; {r^2 \over r^2 + \ell^2} \Big)\,, \quad \quad
a \equiv - {1 \over 2} + {1 \over 2} \sqrt{1 + {L^4 p^2 \over \ell^2}}\,,
\end{eqnarray}
leading to the momentum-space two-point function
\begin{eqnarray}
\langle{\cal O}{\cal O}\rangle \sim p^4 \psi \left( {1 \over 2} + {1 \over 2} \sqrt{1 + {L^4 p^2 \over \ell^2}}\right) \,.
\end{eqnarray}
The digamma function $\psi$ moving out to complex arguments induces a logarithmic branch in the $s\equiv -p^2$ plane starting at $s = \ell^2/L^4$, describing a continuous spectrum of excitations above a mass gap of size $\ell/L^2$. Two-point functions for all the bosonic and fermionic modes of ${\cal N}=8$ gauged supergravity have been calculated in this background \cite{Freedman:1999gk, Bianchi:2000sm, DeWolfe:2012uv}, and all display the same continuous spectrum over a gap. 

\bigskip
\noindent
\underline{\bf ${\cal N}=1$ flow}

\noindent
Consider instead an active scalar $\phi_{\bf 10}(r)$ in the ${\bf 10}$ representation of $SO(6)$, thus dual to a fermionic bilinear Tr $\lambda^2$ in ${\cal N}=4$ SYM, with dimension $\Delta = 3$. There is an RG flow solution \cite{Girardello:1998pd},
\begin{eqnarray}
e^{A(r)} = {r \over L} \left( 1 - {M^2 \over r^2} \right)^{1/2} \,, \quad e^{B(r)} = {L \over r} \,, \quad \phi_{\bf 10}(r) = {\sqrt{3} \over 2} \log \left( r + M \over r-M \right) \,,
\end{eqnarray}
where $M$ is a constant of units length.
This spacetime breaks the symmetry $SO(6) \to SO(3)$ and 32 supercharges to 4 supercharges (${\cal N}=1$ supersymmetry in four dimensions). It may superficially look quite similar to the previous Coulomb branch spacetime, but its interpretation is quite different, as can be seen from the asymptotic expansion of the scalar:
\begin{eqnarray}
\phi_{\bf 10}(r \to \infty) = {\sqrt{3} M \over r} + {
M^3 \over \sqrt{3} r^3} + \ldots \,.
\end{eqnarray}
The leading term indicates there is a source turned on for the operator Tr $\lambda^2$ (as well as the subleading term indicating it has an expectation value). Turning on a constant source for this operator is precisely adding a mass term for the fermion, with mass proportional to $M$. Thus this deformation corresponds not to going to a different state in ${\cal N}=4$ Super-Yang-Mills, but to  modifying the Lagrangian for ${\cal N}=4$ Super-Yang-Mills by adding mass terms, explicitly breaking conformal invariance.\footnote{${\cal N}=1$ supersymmetry is preserved, meaning there must also be a mass term turned on for the six superpartner scalars $X^i$. The corresponding operator has high dimension at strong coupling and is not visible in our truncation, but we can infer it is sourced as well.}

Again we can study the solution to a massless scalar $h$, equivalent to a physical graviton fluctuation. The solution is
\begin{eqnarray}
h = {\alpha^4 \over r^4} \: {}_2F_1 \Big( 2 + {|p|L^2 \over 2 M},  2 - {|p|L^2 \over 2 M}; 2; 1 - {M^2 \over r^2} \Big) \,.
\end{eqnarray}
The correlation function is a little messy but the non-analytic part is
\begin{eqnarray}
\langle {\cal O} {\cal O} \rangle \sim  L^4 p^2 ( L^4 p^2 - 4 M^2) \left(\cdots + \psi \Big( 2 +  {|p|L^2 \over 2 M} \Big) + \psi \Big( 2 -  {|p|L^2 \over 2 M}\Big) \right)\,,
\end{eqnarray}
which thanks to the digamma function has poles at
\begin{eqnarray}
s = {M^2 \over L^4} 4 (n+2)^2 \,.
\end{eqnarray}
Thus unlike the continuous spectrum over a gap of the Coulomb branch flow, here we have excitations only at particular momenta. Since conformal invariance is explicitly broken, we can interpret this as a gauge theory that confines, producing an infinite tower of gauge-invariant ``glueball" states at particular masses. Other fermionic and bosonic modes have been studied, and arrange themselves into similar towers of discrete states, with modes related by the preserved ${\cal N}=1$ SUSY having the same poles \cite{Bianchi:2000sm}.

Thus from these examples we see
\begin{itemize}
	{\item An RG flow geometry with a flowing scalar can correspond either to spontaneously breaking conformal invariance, going to a new state in the CFT, or explicitly breaking conformal invariance, modifying the Lagrangian to something different. The resulting excitations, a continuous spectrum or a discrete set of modes, fit the conformal and confining theories, respectively.}
	{\item Each geometry introduces a characteristic length scale beyond the AdS scale $L$, in these cases $\ell$ or $M$, over which the geometry varies in the radial direction. The existence of this length scale is associated to the breaking of scale invariance, as the spacetime isometry (\ref{GravityScale}) is broken. These distance scales are translated into energy scales $\ell/L^2$ and $M/L^2$ in the dual field theory.}
\end{itemize}

\section{Thermodynamics and AdS/CFT}

We will now describe how the presence of a black hole in AdS space places the dual field theory in a thermal state, and discuss a bottom-up application of such systems to the phase diagram of QCD.

\subsection{Black hole thermodynamics is real thermodynamics}

One interesting thing we can ask about a quantum field theory is, how does the system behave when we give it a temperature $T$? A temperature sets an energy scale, and so correspondingly a length scale as well: conformal invariance is (spontaneously) broken by the state at nonzero temperature. Thus from our understanding of AdS/CFT, we anticipate that in the gravity dual something must sit inside the geometry at a certain radial distance $r_H$ associated to the scale $T$.

In fact, it has been known since Hawking's work in the 70s that there is a metric configuration that has an associated temperature: a black hole. Hawking showed that quantum field theory in the background black hole geometry radiates particle quanta from the horizon as blackbody radiation at the ``Hawking temperature" $T_H$, proportional to the specific gravity at the horizon. In AdS/CFT, a black hole in the geometry at Hawking temperature $T_H$ is dual to the state of the field theory at temperature $T = T_H$.

Since we are interested in configurations in the Poincar\'e patch, we will consider planar black holes (or ``black branes") in anti-de Sitter space, where the horizon at a moment of time is translationally invariant over the $\vec{x}$ coordinates. Such planar black holes are particular cases of the asymptotically AdS metric (\ref{AsymptoticAdS}), where $h(r)$ is the horizon function, whose vanishing defines the location of the horizon $r = r_H$:
\begin{eqnarray}
h(r = r_H)  =0\,.
\end{eqnarray}
In principle, one can determine the Hawking temperature of a given black hole geometry by studying quantum field theory in the curved space background, and making suitable Bogolioubov transformations between in and out states. Once we believe there is a Hawking temperature, however, there is an easier way to determine what it is. A real-time path integral can be turned into a finite-temperature partition function by continuing to Euclidean time $\tau \equiv i t$ and imposing that the imaginary time is periodic,
\begin{eqnarray}
\tau \sim \tau + {1 \over T} \,.
\end{eqnarray}
In flat space, any such periodicity can be chosen. The Euclidean version of a black hole, however, complains unless we pick the periodicity just right. Near the horizon we have $h(r) = h'(r_H) (r - r_H) + \ldots $, and assuming $A(r_H)$ and $B(r_H)$ are not zero or singular, the $\tau/r$ sector of the Euclidanized metric becomes as $r \to r_H$,
\begin{eqnarray}
ds^2 \approx {e^{2B(r_H)}\over h'(r_H)} {dr^2 \over r-r_H} + e^{2A(r_H)} h'(r_H) (r-r_H) d\tau^2 \,.
\end{eqnarray}
Defining a new radial coordinate,
\begin{eqnarray}
\tilde{r} \equiv {2 e^{B(r_H)} \over \sqrt{h'(r_H)}} \sqrt{r-r_H} \,,
\end{eqnarray}
we get
\begin{eqnarray}
ds^2 \approx d\tilde{r}^2 + \tilde{r}^2 {e^{2A(r_H) - 2B(r_H)} h'(r_H)^2 \over 4} d\tau^2 \,.
\end{eqnarray}
If $\tau$ is periodic, this has the structure of a two-dimensional plane in polar coordinates, $ds^2 \approx d\tilde{r}^2 + \tilde{r}^2 d \theta^2$. However, there will be a conical singularity at the origin unless $\theta$ has the proper periodicity, $\theta \sim \theta + 2\pi$. Translating this into a statement about $\tau$, we must have
\begin{eqnarray}
\tau \sim \tau + {4 \pi \over e^{A(r_H)-B(r_H)} h'(r_H)}  \,,
\end{eqnarray}
and thus the Hawking temperature must be,
\begin{eqnarray}
\label{Temp}
T = {e^{A(r_H)-B(r_H)} \over 4 \pi} h'(r_H) \,. 
\end{eqnarray}
Once we have brought the gravity side to Euclidean space at this temperature, we must take the field theory to the same Euclidean periodicity, and so the field theory also lives at temperature $T$.

The thermodynamic variable conjugate to the temperature is the entropy. From the laws of black hole mechanics, it is known that the quantity playing the role of entropy for a black hole is its horizon area divided by four times Newton's constant:
\begin{eqnarray}
S = {A_H \over 4 G_N} \,.
\end{eqnarray}
Since we have a translationally invariant planar black hole, the horizon area and the entropy are infinite. The entropy density per unit $(d-1)$-volume, however, is finite, and can be evaluated as
\begin{eqnarray}
\label{Entropy}
s =  {1 \over 4 G_N} {\int d^{d-1} x \sqrt{g_{d-1}} \over {\rm vol}} = {1 \over 4G_N} e^{(d-1)A(r_H)}  \,.
\end{eqnarray}
Other thermodynamic variables that are often relevant are a chemical potential and charge density for a conserved charge. As we know from the AdS/CFT dictionary, a conserved current in the field theory requires a gauge field in the gravity theory. Consider a conserved $U(1)$ current $J^\mu$ (which could be a part of some larger nonabelian symmetry) and the associated gauge field $A_\mu$. Specializing (\ref{VecFieldExpand}) to the $A_0$ component, the leading term is identified with the source (the chemical potential $\mu$) and the subleading term is the response (the charge density $\rho = \langle J^0 \rangle$),
\begin{eqnarray}
A_0 (r \to \infty) = \mu L  + \ldots - {2 \kappa^2 L^{d-2} \over (d-2)} { \rho \over r^{d-2}} \,.
\end{eqnarray}
A constant term in a gauge field is of course not gauge-invariant; here we have made the assumption that $A_0(r_H) = 0$, in which case identifying $A_0(\infty) = \mu L $ is well-defined.

Let's look at an example spacetime. The (planar) AdS-Schwarzschild black hole has the asymptotic AdS form (\ref{AsymptoticAdS}) with
\begin{eqnarray}
\label{AdSSch}
e^A = e^{-B} = {r \over L} \,, \quad \quad h = 1 - {r_H^d \over r^d}\,.
\end{eqnarray}
The mass of the black hole is proportional to $r_H^d$.
At large $r \gg r_0$, this reverts to AdS$_{d+1}$. The horizon function $h(r)$ has a zero at $r_H$, the location of the horizon.  One can calculate the temperature, 
\begin{eqnarray}
T = {d \over 4 \pi} {r_H \over L^2} \,,
\end{eqnarray}
and entropy density,
\begin{eqnarray}
s = {1 \over 4G} {r_H^{d-1} \over L^{d-1}} \,.
\end{eqnarray}
Once again, we see that the presence of a feature in the geometry at a particular $r_H$ has given us a corresponding energy scale in the field theory --- in this case the temperature ---  at $r_H/L^2$.

For a geometry including a chemical potential and charge density, we can generalize (\ref{AdSSch}) to an AdS-Reissner-Nordstr\"om solution, which is a {\em charged} planar black hole in AdS. We can write it as
\begin{eqnarray}
\label{AdSRN}
e^A = e^{-B} = {r \over L} \,, \quad \quad h = 1 - {r_H^d + Q^2/r_H^{d-2} \over r^d} + {Q^2 \over r^{2d-2}}\,, \quad \quad \quad
A_0 = \mu L \left( 1-  {r_H^{d-2} \over r^{d-2}}\right) \,.
\end{eqnarray}
Here there are two independent parameters, the horizon radius and the charge parameter $Q$. The equations of motion relate the chemical potential $\mu$ and the charge density $\rho$ to $Q$ as
\begin{eqnarray}
\mu = \sqrt{2d-2 \over d-2}{Q \over L^2 r_H^{d-2}}  
  \,, \quad\quad
   \rho ={\sqrt{(d-2)(2d-2)} \over (2 \kappa^2)}{Q \over L^{d-1} }\,,
\end{eqnarray}
while the temperature and entropy can be written
\begin{eqnarray}
T = {d  \over 4 \pi }{r_H \over L^2} \left( 1 - {d-2 \over d} {Q^2 \over r_H^{2d-2}} \right) \,, \quad \quad s = {1 \over 4G} {r_H^{d-1} \over L^{d-1}} \,.
\end{eqnarray}
We may equally well take the two independent parameters to be the temperature $T$ and chemical potential $\mu$.

Thus we see that field theory systems at nonzero temperature and density can be studied by poking and prodding a charged black hole living in AdS space. The same rules we discussed before about calculating correlation functions by varying boundary conditions apply here as well. Much of the milage that has come out of applying the gauge/gravity duality to strongly coupled field theories has come out of these black hole systems.

\begin{figure}
\begin{center}
\includegraphics[scale=0.35]{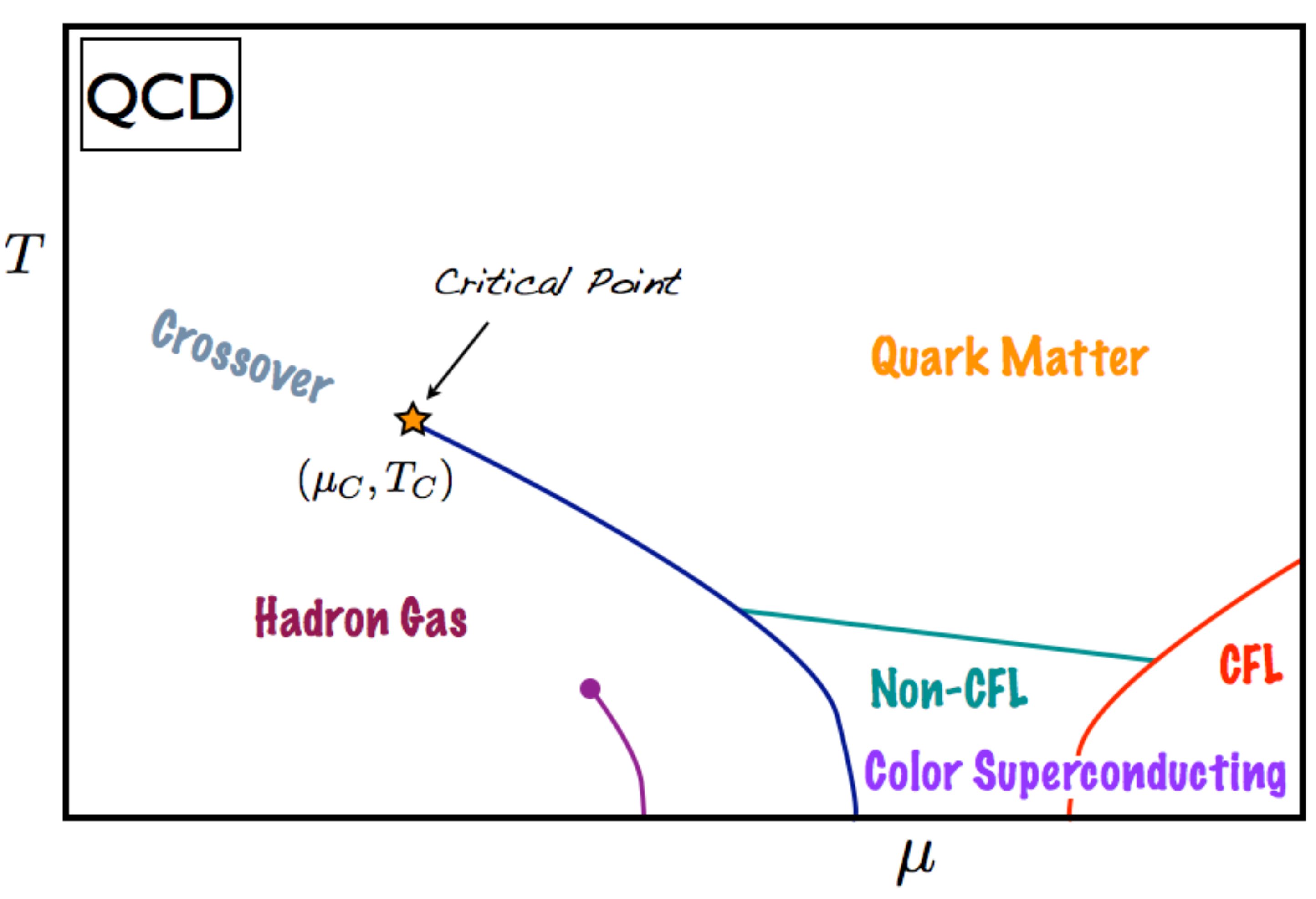}
\caption{A cartoon of the QCD phase diagram. Taken from \cite{DeWolfe:2010he}.
\label{fig:PhaseDiag}}
\end{center}
\end{figure}

\subsection{An example: the phase diagram of QCD}

Quantum chromodynamics (QCD), the theory of the strong nuclear force, is one of the strongly coupled systems we are most interested in understanding. Its Lagrangian might at first seem like just an elaboration on the well-understood form of quantum electrodynamics (QED), but the non-abelian gauge fields change the game entirely. The interactions are strongly coupled at low energies, and quarks and gluons are confined inside hadrons. The large N approach of generalizing to a large number of colors was invented to make QCD more tractable, and large N dynamics of gauge theories find a natural home in the gauge/gravity correspondence. It is natural to ask what we can learn about QCD from this point of view.

One aspect of QCD that is interesting to ask about is its behavior at nonzero temperature and density. Here the density is for the conserved global $U(1)$ baryon number. At zero density, the powerful tool of lattice QCD can be brought to bear, and it is well-known from these studies that varying $T$ from low to high brings us from a region with confinement to a region with deconfinement. In addition, the chiral symmetry (the off-diagonal factor inside the global $SU(N_f) \times SU(N_f)$ symmetry associated to having $N_f$ flavors of quark) goes from being broken to being restored. Because the chiral symmetry is not exact --- it is broken in QCD by non-equal quark masses, along with couplings to electromagnetism and the rest of the Standard Model --- this is not a true phase transition, but a crossover. As the liberation of quarks occurs with increasing temperature, the normalized entropy density $s/T^3$ grows rapidly but smoothly.

It is natural to ask what happens when we introduce a nonzero baryon number density, or alternately a baryon number chemical potential $\mu$. It is generally believed that as $\mu$ increases, the crossover with $T$ should sharpen until at a certain $\mu_C$, it becomes a true phase transition. A line of first-order transitions is expected to appear, terminating at a critical point where the transition is second-order. A cartoon of the QCD phase diagram, including high-density color superconducting and color-flavor-locked phases we will not discuss, appears in figure~\ref{fig:PhaseDiag}.

However, it turns out it is very hard to study $\mu \neq 0$ in lattice QCD. The chemical potential in Euclidean space introduces complex terms in the action, which are much harder to deal with --- the so-called ``sign problem". AdS/CFT, on the other hand, doesn't have a sign problem. It is perfectly happy to study nonzero $\mu$ --- all one has to do is turn on a gauge field in the presence of the AdS black hole. So it is natural to ask what the gauge/gravity correspondence can tell us.

Of course, AdS/CFT is not a perfect probe either. In particular, we have no gravity dual for QCD itself, not even QCD with a large number of colors.  The signature dualities of the gauge/gravity correspondence all involve conformal field theories, and QCD is not conformal. As we have mentioned, turning on certain scalar fields in the gravity background can break conformal invariance by turning on sources for field theory operators, but it is not known how to get to QCD precisely in a top-down fashion.

So in this section, we will explore what a bottom-up model can do. Rather than trying to derive QCD from first principles, we will generate a recipe for a holographic theory that has the QCD properties we particularly want to explore --- namely its thermodynamics. Then we will cook up the recipe and see how it tastes.

Since we are building a bottom-up model, we can have whatever gravity fields we want. We will pick three:

\begin{itemize}
	{\item The metric $g_{\mu\nu}$, dual to the energy and momentum $T_{\mu\nu}$.}
	{\item A $U(1)$ gauge field $A_\mu$, dual to the conserved current of baryon number $J^\mu$.}
	{\item An almost-massless scalar field $\phi$.} 
\end{itemize}

The scalar field requires a little more explanation: it is there to model the running of the QCD coupling. In type IIB supergravity on $AdS_5 \times S^5$, there is a massless scalar dual to the ${\cal N}=4$ Lagrangian, for which turning on a source is just shifting the exactly marginal ${\cal N}=4$ SYM coupling constant. The QCD coupling, on the other hand, is not exactly marginal, but runs, though at higher energy it runs slowly. We introduce a scalar that is almost massless, so when we turn it on in the gravity background, it sources an almost marginal operator and introduces a slowly running coupling in the field theory. This is a very ``bottom-up" move --- we are not exactly reproducing the QCD coupling, but by introducing something that runs slowly, we hope to imitate its properties.

Given these fields, we have a gravity Lagrangian
\begin{eqnarray}
\label{QCDPhaseL}
{\cal L} = R - {1 \over 2} (\partial \phi)^2 - V(\phi) - {1 \over 4} f(\phi) F_{\mu\nu} F^{\mu\nu}\,,
\end{eqnarray}
for some potential $V(\phi)$ and gauge kinetic function $f(\phi)$. In principle these could be whatever we wanted, and each choice would define a different (unknown) strongly coupled dual field theory. What we will do is to choose these functions to match known lattice data, and then step off to $\mu \neq 0$ where lattice data has a hard time following.

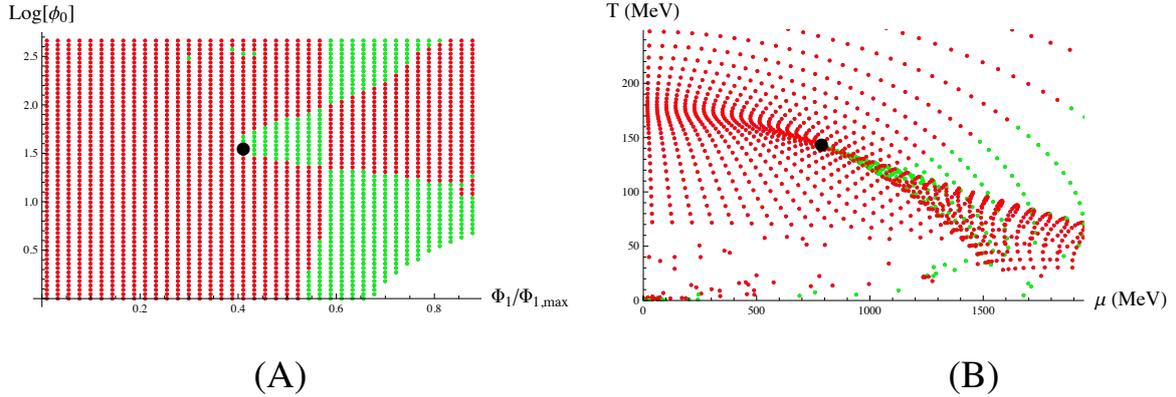
\begin{figure}
  \centerline{\includegraphics[width=6in]{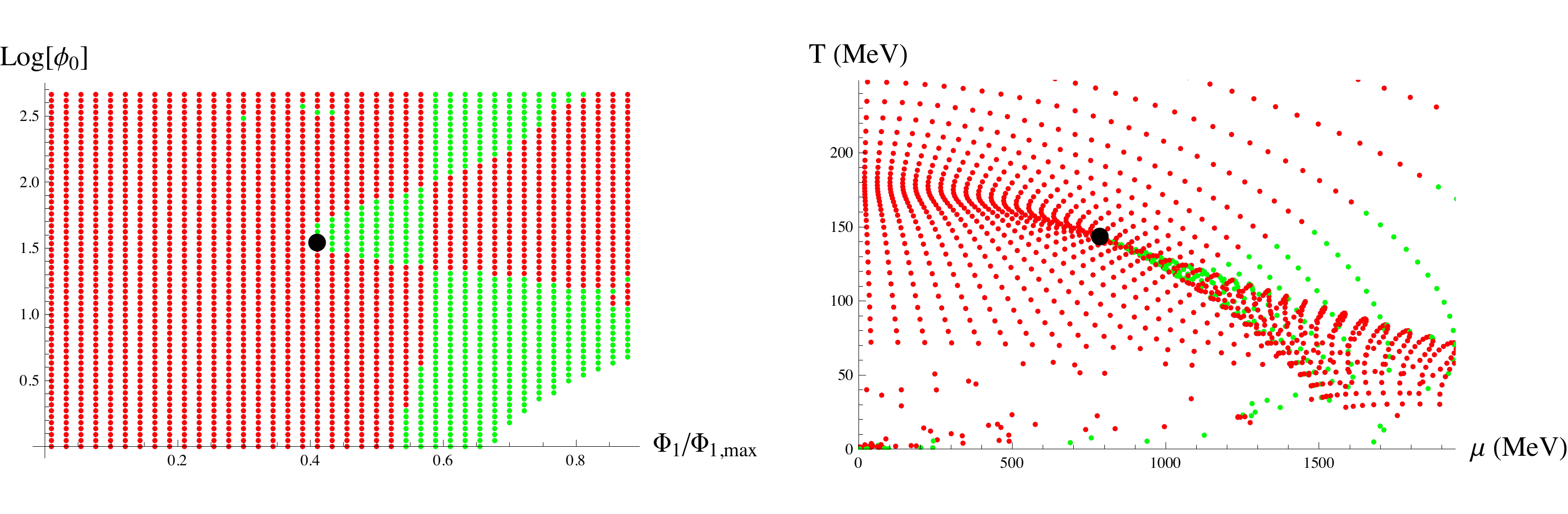}}\begin{picture}(0,0)(0,0)\put(90,0){\Large (A)}\put(350,0){\Large (B)}\end{picture}
  \caption{Numerically generated black holes.  Each dot represents a numerically generated solution.  Red points are thermodynamically stable, while green points are thermodynamically unstable. Taken from \cite{DeWolfe:2010he}.}
 \end{figure}

Lattice data makes predictions for the entropy density $s(T)$ and the quark suceptability $\chi(T)\equiv {\partial \rho \over \partial \mu}(T)$. Making a choice of functions \cite{Gubser:2008ny, Gubser:2008yx, DeWolfe:2010he}
\begin{eqnarray}
\label{VChoice}
  V(\phi) =  {-12 \cosh \gamma\phi + b\phi^2 \over L^2}  \,, \quad\quad f(\phi) = {\rm{sech} \left[ {6 \over 5} (\phi-2) \right] \over \rm{sech} {12 \over 5}} \,,
\end{eqnarray}
with $\gamma = 0.606$ and $b = 2.057$, we can numerically generate a series of (uncharged) black hole solutions to the theory. These solutions are like generalizations of the AdS-Schwarzschild solution (\ref{AdSSch}), but with the scalar field turned on as well. Like AdS-Schwarzschild, the black hole geometries have a horizon and an associated temperature and entropy. Using (\ref{Temp}) and (\ref{Entropy}), we can determine the thermodynamics of the field theory states dual to these black holes. Although $\rho$ and $\mu$ vanish for uncharged black holes, it is possible to derive an expression for their derivative. Putting these together, thanks to the choices (\ref{VChoice}), this ensemble of black holes have thermodynamic properties matching well the predictions of lattice QCD. In particular, we have ``baked in" the crossover. Note that (\ref{VChoice}) is not in any sense a perfectly optimized solution in the space of all potentials; it is one choice that works pretty well.

Having allowed lattice thermodynamics to fix the model (\ref{QCDPhaseL}), (\ref{VChoice}) at $\mu=0$, we proceed to use the model to generate charged black holes with $\mu \neq 0$. In practice, this is done by seeding values of $\phi$ and $dA_0/dr$ at the horizon and numerically solving the Einstein-Maxwell-dilaton equations of motion. One arrives at generalizations of the AdS-Reissner-Nordstr\"om solutions (\ref{AdSRN}), again with a nonzero scalar added. This ensemble of black holes can be analyzed for thermodynamics, to determine $s(T, \mu)$ and $\rho(T, \mu)$.

The result is that AdS/CFT ``knows" that the crossover is supposed to sharpen into a first-order phase transition, and the ensemble of black holes have thermodynamics displaying this behavior. For example, $\rho(\mu)$ at $T > T_C$ displays crossover behavior, but at $T< T_C$, it instead becomes multi-valued, bending back on itself. This multivalued behavior is the hallmark of a first-order phase transition; the two positive-slope solutions at a given $\mu$ are the two competing phases, while the negative-slope solution is thermodynamically unstable. Which of the two stable solutions is physically realized is resolved by determining which minimizes the free energy, and given the definition of the correspondence, the free energy is just the classical gravity action evaluated on each solution. The locus on the phase diagram where the two solutions have exactly equal free energies is the first-order line.

\begin{figure}
\begin{center}
\includegraphics[scale=0.54]{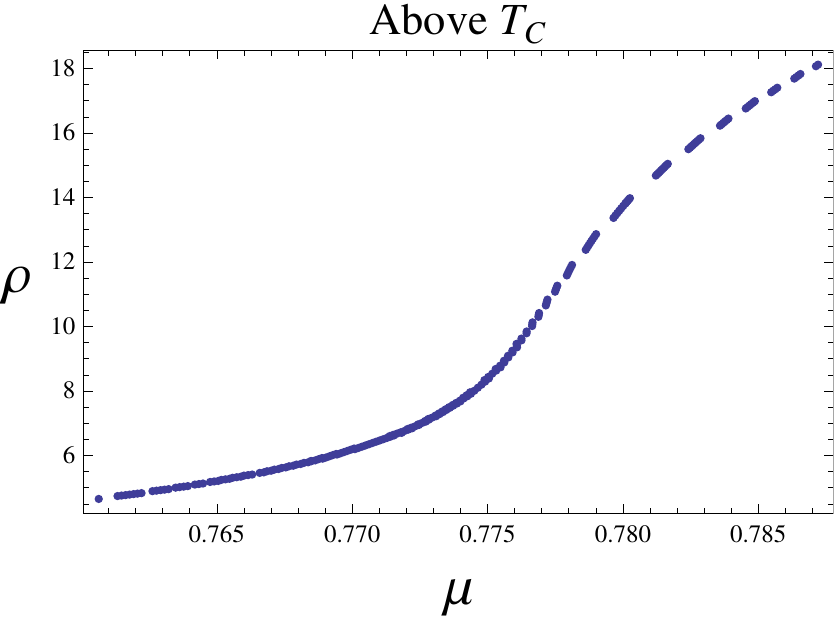}
\includegraphics[scale=0.46]{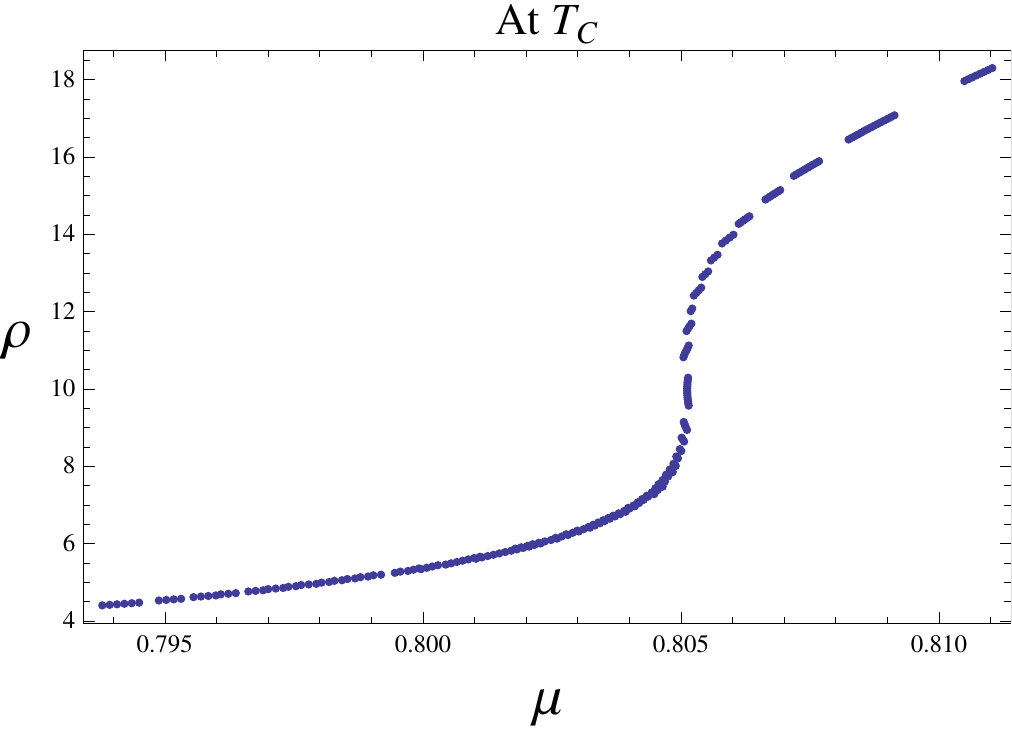}
\includegraphics[scale=0.46]{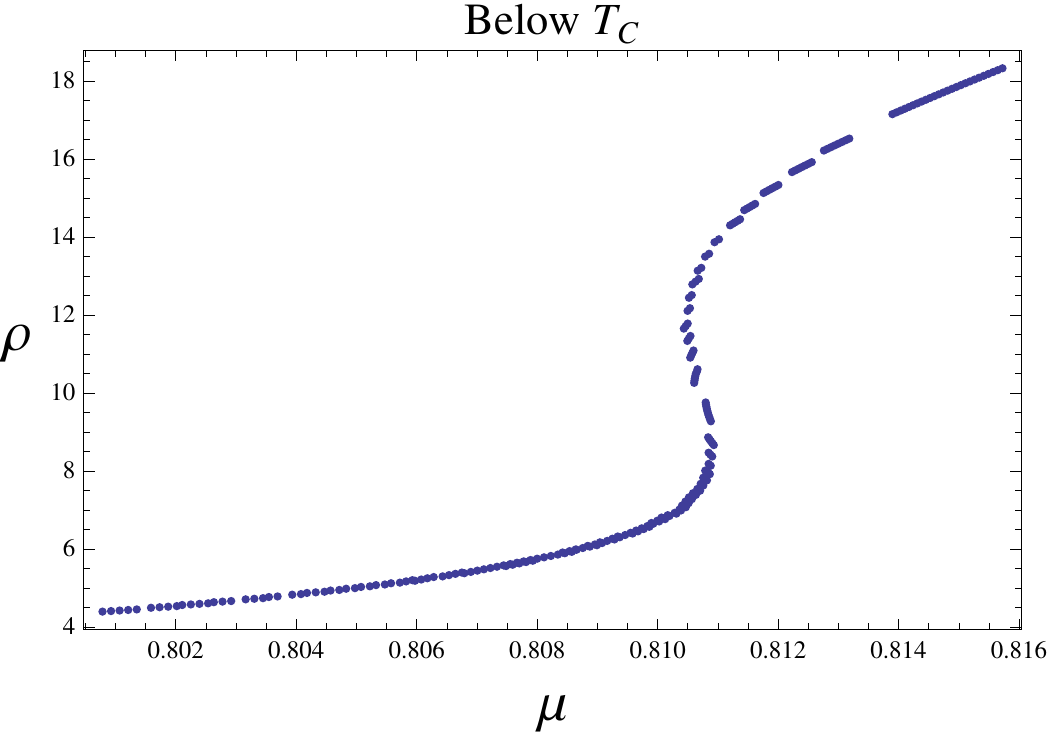}
\caption{The baryon density $\rho$ as a function of chemical potential $\mu$ for several values of $T$ near the critical point. Taken from \cite{DeWolfe:2010he}.
\label{fig:rhoMu}}
\end{center}
\end{figure}

One can then ask about the end of the first-order line, which should be a critical point. Critical points can be classified by the critical exponents by which susceptibilities diverge as it is approached. The QCD critical point is generally believed to be in the same universality class (sharing the same exponents) as the 3D Ising model, which also matches an ordinary liquid/gas critical point. Does AdS/CFT know about the critical exponents?

The answer is a nice illustration of how much AdS/CFT ``knows", and what it doesn't know. The susceptibilities of the black hole ensemble do diverge near the critical point, with exponents that match the 3D Ising model --- but in the mean-field approximation. The corrections to the exponents from quantum fluctuations are invisible to the gravity theory \cite{DeWolfe:2010he}.

So on the one hand, it is remarkable that an ensemble of black hole geometries, designed to imitate the known thermodynamics of QCD at zero density, ``knows" to sharpen the crossover into a line of phase transitions as the chemical potential increases, with a critical point in the anticipated universality class to boot. But the critical point is only in the mean-field limit. Presumably, the same large N, large coupling approximations that suppress quantum corrections on the gravity side and allow us to do classical GR, also suppress the quantum corrections to the critical exponents. If we could include quantum corrections to the gravity theory, the full exponents might reveal themselves, but as it is we are working in a semiclassical limit where fluctuations are suppressed. Still, these bottom-up models make the prediction that if a theory with a crossover like QCD is extended to nonzero density, a critical point should indeed appear.

\section{Real-time Correlators and the Shear Viscosity of the Quark-Gluon Plasma}

At low temperatures, QCD states are hadrons. As temperature increases, the theory moves through the crossover and is expected to eventually reach a state of liberated quarks and gluons. This is consistent with our understanding of asymptotic freedom, which tells us non-abelian gauge theories like QCD are weakly coupled at high energy. Such a phase of mostly-free quarks and gluons was dubbed ``the quark-gluon plasma" (QGP). Two decades ago it was expected that heavy ion collisions like those then being planned at the Relativistic Heavy Ion Collider (RHIC) would create a state of hadronic matter with an effective high temperature that would produce this quark-gluon plasma.

When RHIC began collecting data for gold-gold collisions (and later LHC for lead-lead collisions), however, this was not what was observed. A state of matter consistent with almost free color charged particles was not observed. Instead, the hadronic matter behaved in a way consistent with being a fluid, with no quasiparticle description at all. This state of matter --- still called the quark-gluon plasma --- quickly freezes out into hadrons, but in the meantime seems to flow with a viscosity that is extremely small compared to its entropy density. The ratio of shear viscosity to entropy density for water is on the order 2; for liquid helium, 0.7. The QGP shows up at $0.12 \pm 0.08$, as low as any substance ever seen (for reviews in our context see \cite{CasalderreySolana:2011us, Adams:2012th, DeWolfe:2013cua}).

This observation seems to cry out for AdS/CFT to address it. The gauge/gravity duality specializes in strongly coupled gauge theories with no quasiparticle description. And in fact, a theoretical suggestion that the QGP might have such a low viscosity had emerged several years before the RHIC data came in.

\subsection{Hydrodynamics and transport}

In equilibrium, a substance can be characterized by its temperature $T$, as well as a chemical potential $\mu$ for any conserved current. Hydrodynamics describes a fluid close enough to equilibrium that a notion of a temperature can be assumed to hold locally: we assume there exists $T(x)$ (and $\mu(x)$).

The dynamics of energy and momentum are captured by the energy-momentum tensor, which can be written near equilibrium in a derivative expansion. One may write this in a fully covariant form, but for us it is enough to choose a local rest frame where $T_{0i} = 0$, where $i$ runs over the $d-1$ spatial coordinates. Then to first order in derivatives, we have
\begin{eqnarray}
\label{FluidEMT}
T_{00}&=& \varepsilon \,, \\
T_{ij}&=& \delta_{ij} p - \eta \Big( \partial_i u_j - \partial_j u_i - {2 \over d-1} \delta_{ij} \partial_k u^k \Big) - \xi \delta_{ij} \partial_k u^k \,.
\end{eqnarray}
Here $\varepsilon$ and $p$ are the energy density and pressure, characterizing the fluid at leading order, and $u_i$ is the fluid velocity vector. The parameters characterizing the terms first-order in derivatives are called {\em transport coefficients}: the shear viscosity $\eta$ and the bulk viscosity $\xi$. If a conserved current is present, the expansion of $J^\mu$ contains another transport coefficient, the conductivity.

Below we will be interested in calculating the shear viscosity. This can be done using the theory of linear response: if we turn on a source, how does the system respond? Imagine adding a source $g^{\mu\nu}$ for the energy-momentum tensor as a perturbation to the Hamiltonian,
\begin{eqnarray}
H = H_0 + H' = H_0 + \int d^{d-1}x \, g^{\mu\nu} T_{\mu\nu}\,.
\end{eqnarray}
The time rate of change of $\langle T_{\mu\nu}\rangle$ is then given in terms of a commutator,
\begin{eqnarray}
{d \langle T_{\mu\nu} \rangle \over d t} = i \langle [ H, T_{\mu\nu} ]\rangle \,,
\end{eqnarray}
and integrating we find
\begin{eqnarray}
\delta \langle T_{\mu\nu}(\vec{x},t)\rangle &=& i \int_{t_0}^t dt' \, \Big\langle \Big[ \int d^{d-1}x' g^{\rho\sigma}(\vec{x}',t') T_{\rho\sigma}(\vec{x}',t'), T_{\mu\nu} (\vec{x},t)\Big]\Big\rangle\\
&=& \int d^dx' g^{\rho\sigma}(\vec{x}',t') G^R_{\rho\sigma,\mu\nu}(\vec{x}', t,;\vec{x},t) \,,
\end{eqnarray}
where we defined the retarded Green's function,
\begin{eqnarray}
G^R_{\rho\sigma,\mu\nu}(\vec{x}', t,;\vec{x},t) = i \theta(t-t')[ T_{\rho\sigma}(\vec{x}',t'), T_{\mu\nu} (\vec{x},t)] \,,
\end{eqnarray}
whose $\theta$-function encodes the fact that the response must come after the source that provokes it. In momentum space, this becomes
\begin{eqnarray}
\delta \langle T_{\mu\nu} (\omega, \vec{k}) \rangle = g^{\rho\sigma}(\omega, \vec{k}) G^R_{\rho\sigma;\mu\nu}(\omega, \vec{k}) \,.
\end{eqnarray}
Thus the information on the response of the energy-momentum tensor to a perturbing source is contained in the retarded Green's function.

We can determine the shear viscosity from a transverse traceless mode of the Green's function for energy-momentum tensor in the zero momentum limit, for example $G^R(\omega) \equiv G^R_{xy;xy}(\omega, \vec{k}=0)$. Since the shear viscosity arises at first order in derivatives in (\ref{FluidEMT}), it will appear with $\omega$ as its coefficient. We can thus extract it using a so-called Kubo formula:
\begin{eqnarray}
\label{Kubo}
\eta = - \lim_{\omega \to 0} {1 \over \omega} G^R(\omega) \,.
\end{eqnarray}
Thus to calculate this transport coefficient in AdS/CFT, we need to calculate the retarded propagator $G^R$.

\subsection{Real-time correlators in AdS/CFT}
\label{RealTimeSec}

In Euclidean AdS/CFT, including the RG flows we considered, the two solutions to fluctuation equations far from the boundary take the form of one regular, one divergent, and the AdS/CFT prescription is to choose the regular boundary condition.

Black hole horizons in Lorentzian signature are different. There the typical solutions of linearized equations of motion near the black hole horizon $r_H$ take the form,
\begin{eqnarray}
\label{InfallingOutgoing}
X(r,t, \vec{x})_{\omega;\:\vec{k} = 0} &=& e^{-i \omega t}x_0 (r -r_H)^{\pm i \alpha \omega} (1 + \ldots)\\
&=& x_0 e^{i \omega ( \pm \alpha \log (r-r_H) - t)} \,,
\end{eqnarray}
where $\alpha$ is a constant. These solutions can be identified as infalling and outgoing modes. What is the boundary condition to choose?

The prescription of Son and Starinets \cite{Son:2002sd} is that keeping the {\em infalling} modes only, corresponds to calculating the {\em retarded} Green's function. The reasoning behind this choice is that it corresponds to the dissipation of linear response: we poke the system with the black hole and let it relax, and watch the excited modes fall behind the horizon. The fact that the information of the modes is lost into the black hole corresponds to the dissipative process; it is physical for modes to fall into the horizon, just as it is physical that response follow cause, as in the retarded propagator. Modes coming purely out of the horizon would correspond to the causality-reversed case of the advanced propagator. (One can make these arguments more rigorous; see \cite{Skenderis:2008dg}).

Thus to calculate a retarded Green's function at finite temperature, we solve the linearized fluctuation equation in the Lorentzian signature black hole geometry, and impose {\em infalling} boundary conditions (the lower sign in (\ref{InfallingOutgoing})).

\subsection{Calculating the shear viscosity to entropy density }

The AdS/CFT result for the ratio of shear viscosity to entropy density  was originally done by \cite{Kovtun:2004de} and can be calculated in a number of ways. Here we follow the calculation of \cite{Gubser:2008sz}. We specialize to a four-dimensional field theory with gravity dual involving an ordinary Einstein gravity, but do not make any further restriction on the theory.

To calculate the retarded Green's function $G^R_{xy;xy}$, we will solve the fluctuation equation for the graviton mode $h_{xy} \equiv Z$ with $\omega \neq 0, \vec{k}=0$. As mentioned before, a transverse traceless graviton mode obeys the Klein-Gordon equation of a massless scalar, which in the asymptotically AdS background of (\ref{AsymptoticAdS}) takes the form
\begin{eqnarray}
\label{ZKG}
Z'' + (4A' - B'+ {h'\over h}) Z' + \omega^2 {e^{2B-2A} \over h^2} Z =0\,.
\end{eqnarray}
Let us define the coefficient of the $Z'$ term as $p(r)$. The theory of ordinary differential equations then tells us that the quantity
\begin{eqnarray}
{\cal F} \equiv e^{\int p(r) dr} \times{\rm Wronskian} \,,
\end{eqnarray}
is independent of $r$. Choosing the solutions $Z$ and $Z^*$ to put into the Wronskian, we obtain
\begin{eqnarray}
{\cal F}= e^{4A-B} h \, {\rm Im} (Z^* Z') \,.
\end{eqnarray}
We will make our progress by evaluating this quantity in the UV (near the boundary) and in the IR (near the horizon).

Near the boundary, the solution to (\ref{ZKG}) takes the form,
\begin{eqnarray}
\label{ZExpansion}
Z(r \to \infty) = Z^{(0)}_{\rm UV} + \ldots + {Z^{(4)}L^8 \over r^4} + \ldots \,,
\end{eqnarray}
and with $Z^{(0)}_{\rm UV}$ identified as the source, holographic renormalization identifies the expectation value for the dual operator and the resulting Green's function are 
\begin{eqnarray}
\langle {\cal O} \rangle = - 4 {L^3 \over 2 \kappa^2}  Z^{(4)}
\quad \quad \to \quad \quad
G_R = - 4 {L^3 \over 2 \kappa^2}  {Z^{(4)} \over Z^{(0)}_{\rm UV}} \,.
\end{eqnarray}
Near the boundary we also have
\begin{eqnarray}
e^{4A-B} h \to {r^5 \over L^5} \,.
\end{eqnarray}
In calculating the Im$(Z^* Z')$ part of ${\cal F}$, the coefficients of the unwritten middle terms in (\ref{ZExpansion}) are proportional to $Z^{(0)}_{\rm UV}$, and thus the corresponding cross terms are all real; the first non-real term involves $Z^{(0)*}_{\rm UV} Z^{(4)}$. Choosing a normalization where $Z^{(0)}_{\rm UV} = 1$, this is just proportional to the imaginary part of the retarded Green's function,
\begin{eqnarray}
\label{ImagGreen}
{\rm Im} \, G^R = {1 \over 2 \kappa^2} {\cal F} \,.
\end{eqnarray}
Now turn to the IR. Near the horizon we have
\begin{eqnarray}
Z &=& Z^{(0)}_{\rm IR} (r - r_H)^{-i \alpha \omega} + \ldots\\
&=& Z^{(0)}_{\rm IR} ( 1 - i \alpha \omega \log (r-r_H) + \ldots) \,,
\end{eqnarray}
where
\begin{eqnarray}
\alpha = {e^{B-A} \over h'}\Bigg|_{r=r_H} \,,
\end{eqnarray}
which implies
\begin{eqnarray}
{\cal F}= - e^{3A(r_H)} \omega |Z^{(0)}_{\rm IR}|^2 \,.
\end{eqnarray}
Plugging into (\ref{ImagGreen}) and the Kubo formula (\ref{Kubo}), we find
\begin{eqnarray}
\eta = {e^{3A(r_H)} \over 16 \pi G_N} |Z^{(0)}_{\rm IR}|^2 \,,
\end{eqnarray}
where we used $2 \kappa^2 = 16 \pi G_N$. We see this is proportional to the gravitational expression for the entropy density in terms of the horizon area (\ref{Entropy}). Thus we have
\begin{eqnarray}
{\eta \over s} = {1 \over 4\pi} |Z^{(0)}_{\rm IR}|^2 \,.
\end{eqnarray}
To complete the calculation, we have to relate $Z^{(0)}_{\rm IR}$ to $Z^{(0)}_{\rm UV}$. In general one would not be able to do this without solving the fluctuation equation everywhere. However, here we have a trick up our sleeve.  In the $\omega \to 0$ limit, the Klein-Gordon equation (\ref{ZKG}) becomes
\begin{eqnarray}
\partial_r (\log Z')  + \partial_r (4A - B + \log h) = 0\,.
\end{eqnarray}
Integrating this, we obtain
\begin{eqnarray}
Z(r) = {\rm const} + {\rm const}' \int_r^\infty dr' {e^{-4A+B} \over h} (r')\,.
\end{eqnarray}
Going to the UV $r \to \infty$, we find
\begin{eqnarray}
{\rm const} = Z^{(0)}_{\rm UV} \,,
\end{eqnarray}
while going to the IR where $h \propto r-r_H$,
\begin{eqnarray}
Z = {\rm const} + {\rm const}'' \log (r-r_H) \,,
\end{eqnarray}
and thus const $= Z^{(0)}_{\rm IR}$.  Putting these together we have
\begin{eqnarray}
Z^{(0)}_{\rm IR} = Z^{(0)}_{\rm UV} = 1\,,
\end{eqnarray}
and thus 
\begin{eqnarray}
{\eta \over s} = {1 \over 4 \pi} \,.
\end{eqnarray}
The particularly simple form of the fluctuation equation in the low-energy limit allowed us to relate the UV and IR directly, and solve for the shear viscosity to entropy density ratio. (We note that for other transport coefficients, such a simplification does not occur.) Thus it is {\em universally} the case for any theory with an Einstein gravity dual that this ratio will obtain.

This result represents another way to proceed given that we don't have an exact gravity dual of QCD: look for general features of large N gauge theories that span a wide class of cases. We do not have a gravity dual for QCD, but we now have seen that it is a generic feature of large-N, strongly coupled field theories that their viscosity to entropy density ratio is very low; so it is not unreasonable to expect that QCD might have this property as well, and indeed experimentally this turns out to be the case.

\section{Holographic Superconductors}

So far we have encountered systems with scalars turned on, leading to RG flow geometries, and systems with gauge fields turned on, leading to a density and chemical potential for the corresponding dual conserved charge. It is also possible to turn on scalars and gauge fields simultaneously. If the scalars are neutral with respect to the gauge field, these geometries are qualitatively similar to ones we have already studied; in fact the QCD phase diagram spacetimes we considered are examples of this class.

If a charged scalar is turned on, however, things are qualitatively different. The corresponding $U(1)$ is now broken. Such geometries are generally referred to as {\em holographic superconductors} \cite{Gubser:2008px, Hartnoll:2008vx, Hartnoll:2008kx, Gubser:2008pf}, because of the spontaneous breaking of the symmetry. Probably a better name would be {\em holographic superfluids}, because the dual field theory current $J^\mu$ is global, not gauged. But people like to imagine it would be easy to weakly gauge the current, and the holographic superconductor name has stuck.

Consider a gravity action of the form
\begin{eqnarray}
\label{ChargedScalarL}
S_{\rm grav} = {1 \over 2 \kappa^2} \int d^{d+1}x \left( R - {1 \over 4} F_{\mu\nu} F^{\mu\nu} - |(\partial_\mu - i e A_\mu) \phi|^2 - V(\phi) \right)\,,
\end{eqnarray}
containing a charged scalar $\phi$. We can imagine the field $\phi$ is dual to the ``Cooper pair", the charged bosonic composite whose condensation leads to superconductivity. 

One class of solutions to (\ref{ChargedScalarL}) that always exists is the charged black hole, the AdS-Reissner-Nordstr\"om solution (\ref{AdSRN}) with vanishing scalar $\phi = 0$. One can choose any $T$ and $\mu$ for these solutions, and the $U(1)$ is unbroken. Since the dual theory is conformal, only the ratio $T/\mu$ really matters, and black holes with the same ratio will be coordinate-equivalent under the scale transformation.

However, it turns out that other solutions exist for certain values of $T$ and $\mu$. To anticipate why this might be the case, let's look at some interesting properties of the AdSRN solution at zero temperature. If we choose parameters $Q^2=d/(d-2) r_H^{2d-2}$, we get $T=0$ but $\mu >0$, and the horizon function becomes
\begin{eqnarray}
h  = 1 - {2d-2 \over d-2} {r_H^d \over r^d} + {d \over d-2} {r_H^{2d-2} \over r^{2d-2}} = {d (d-1) \over r_H^2} (r-r_H)^2 + {\cal O}((r-r_H)^3)\,.
\end{eqnarray}
The geometry still has a horizon, but there is now a double zero in the horizon function; this is the signature of an {\em extremal} black hole, which has the minimum possible mass for a given charge. Near the horizon, the extremal metric takes the form
\begin{eqnarray}
ds^2 = - {(r-r_H)^2 \over L_2^2} dt^2 + {L_2^2 d(r-r_H)^2 \over (r-r_H)^2} &+& {r_H^2 \over L^2} d\vec{x}^2 \,, \quad \quad A_0 = {\sqrt{2}\over L_2}  (r-r_H ) \,, \\ L_2^2 &\equiv& {L^2 \over d(d-1)} 
 \,. 
\end{eqnarray}
which is AdS$_2 \times \mathbb{R}^{d-1}$, with the time and radial directions combining into AdS$_2$ with characteristic length $L_2$, and a constant electric field in the radial direction.

Consider fluctuations of the charged scalar in this background. The scalar receives an additional effective contribution to its mass from the coupling to the electric field, which near the horizon takes a simple form:
\begin{eqnarray}
\label{meff}
m^2_{\rm eff} = m^2 + e^2 g^{tt} A_0^2 = m^2 - 2 e^2 \,.
\end{eqnarray}
Now an interesting thing can occur: even if the mass $m^2$ satisfies the BF bound in AdS$_{d+1}$, it can be that the effective mass $m_{\rm eff}^2$ may violate the AdS$_2$ BF bound,
\begin{eqnarray}
\label{AdS2BF}
m_{\rm eff}^2 L_2^2 \geq - {1 \over 4} \,,
\end{eqnarray}
if the electric charge $e$ of the scalar field is sufficiently strong.
Such a violation suggests that while the scalar field may be stable in the AdS vacuum, it develops an instability  near the horizon of a charged black hole. We might expect the scalar to condense, forming a condensate around the black hole. For $T$ nonzero but small, the near-horizon geometry is not  precisely AdS$_2$, but the charge contribution to $m^2_{\rm eff}$ can be seen to get smaller as $T$ is increased, suggesting that this condensation phenomenon will be strongest at low temperatures.

We have now motivated that there might be asymptotically AdS charged black hole solutions with a nonzero scalar field turned on, preferentially at lower temperatures. We can look for such solutions, imposing the choice that the scalar profile is only associated to the dual operator having an expectation value turned on, but no source; this means the solutions will be dual to states in the {\em same} dual field theory as the AdSRN backgrounds.

\begin{figure}
\begin{center}
\includegraphics[scale=0.4]{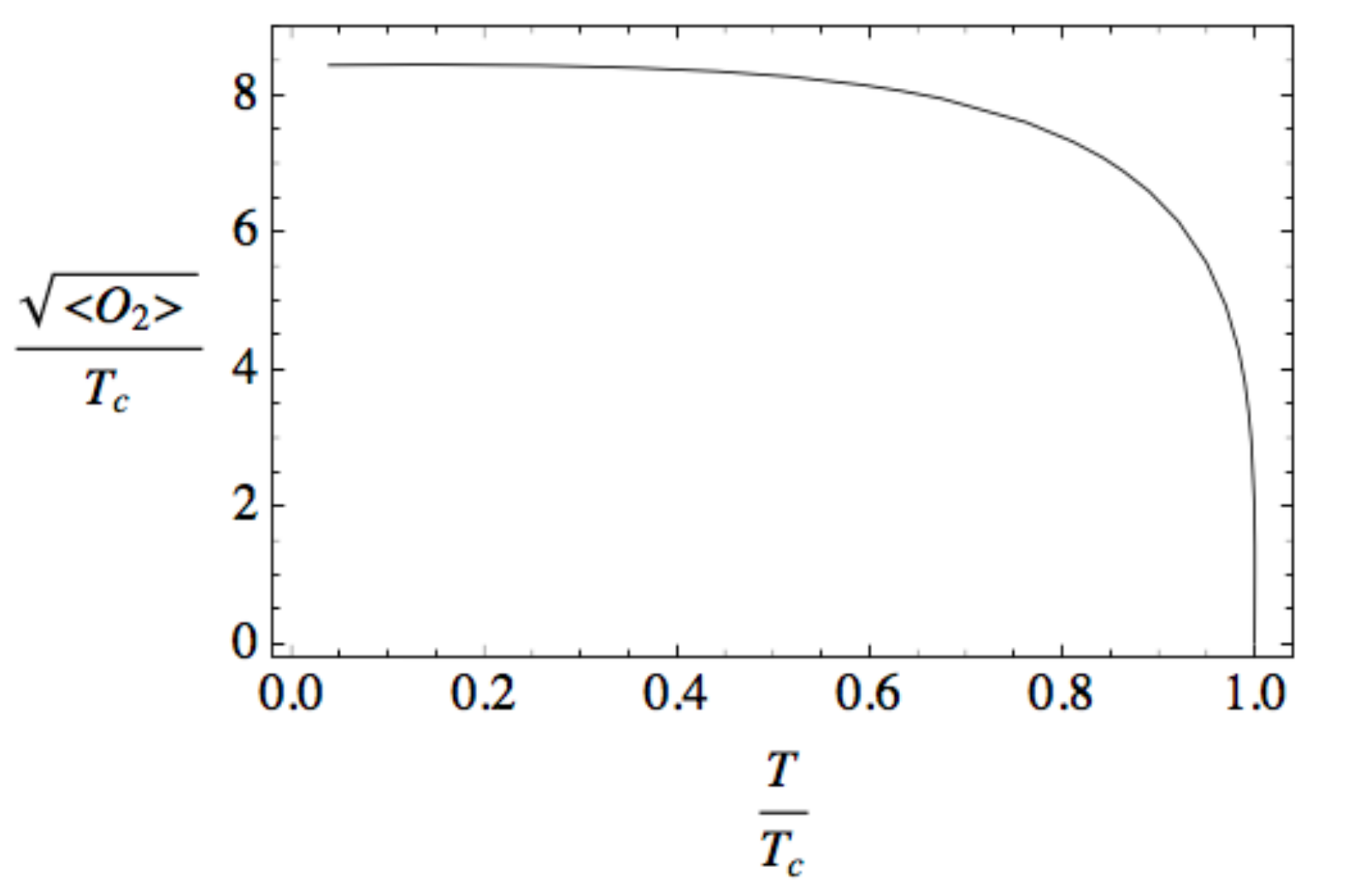}
\includegraphics[scale=0.4]{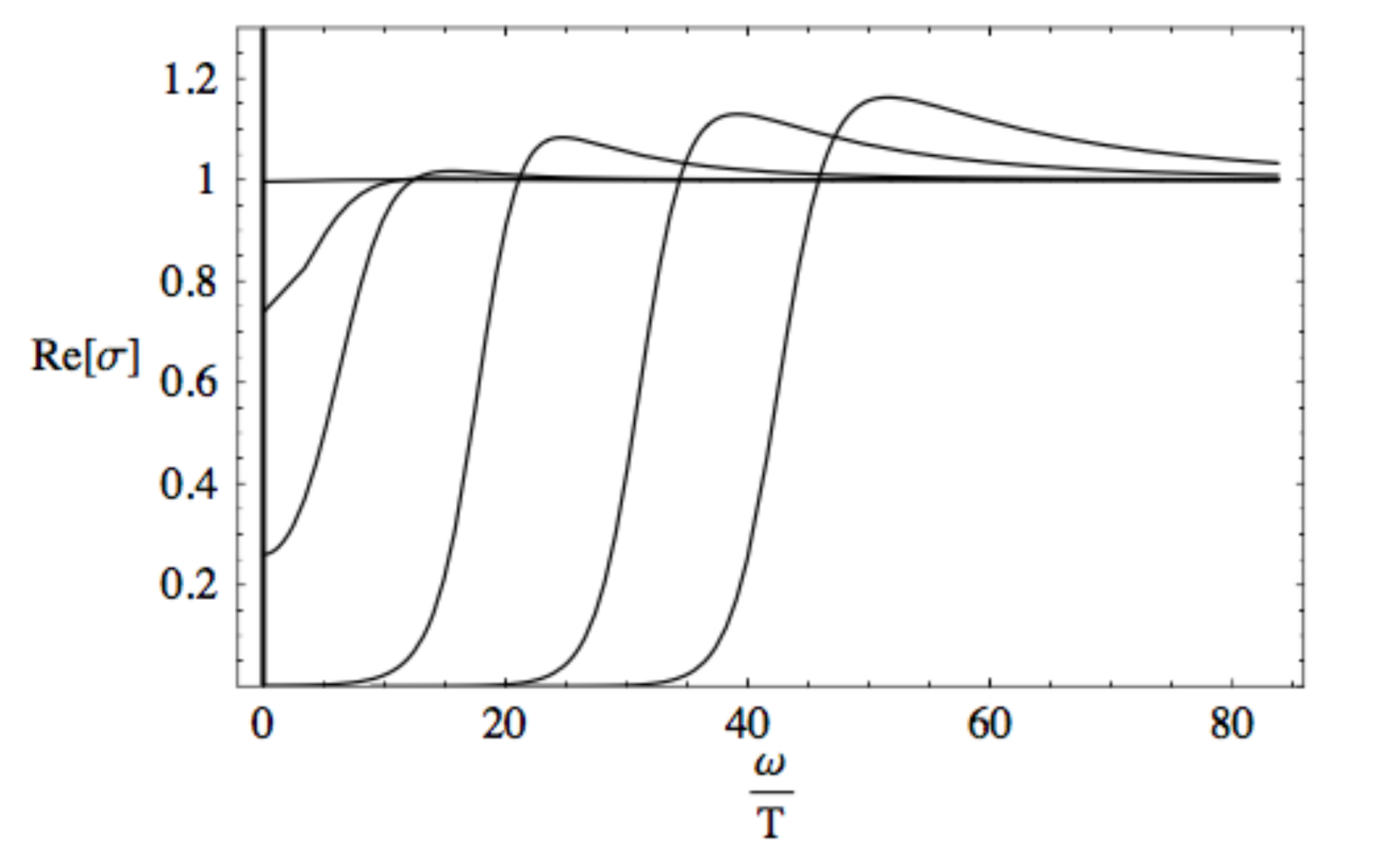}
\caption{The value of the condensate $\langle {\cal O}_2\rangle$ compared to the critical temperature $T_c$ as temperature is varied, with $d=3$, $m^2 L^2 = -2$ and $e=1$, from \cite{Hartnoll:2008vx}.
\label{fig:HoloSup}}
\end{center}
\end{figure}

It turns out these solutions do exist, and furthermore exist only below a certain critical temperature $T_c$, whose magnitude is set by the only other parameter associated to a mass scale, the chemical potential: $T_c \sim \mu$. As the temperature is decreased in this family of solutions, the value of the condensate $\langle {\cal O}\rangle$ grows. For example, consider the case $d=3$ with the scalar mass $m^2 L^2 = -2$, which shows up in ${\cal N}=8$ gauged supergravity and was our example when we went over holographic renormalization, and let it be in the regular quantization so the dual operator ${\cal O}_2$ has dimension 2. One finds solutions with asymptotic scalar field
\begin{eqnarray}
\phi(r\to  \infty) = {0 \over r} + {{\cal O}_2 \over r^2} + \ldots \,,
\end{eqnarray}
with the condensate as a function of temperature given in figure~(\ref{fig:HoloSup}a) \cite{Hartnoll:2008vx}. 

Thus there are two solutions to the same theory with the same $T/\mu$, one with broken $U(1)$ symmetry and one with unbroken. As in the QCD phase diagram discussion, this corresponds to two different states at the same point in the phase diagram. Which solution is actually preferred is a question of which minimizes the free energy, which as discussed previously is the (fully renormalized) classical gravitational action. Calculating this, one indeed finds that the solutions with the condensate are thermodynamically preferred. Thus at high temperatures, only AdSRN solutions exist and the dual field theory is in a regular phase, but as the temperature is lowered below $T_c$, the holographic superconductor solutions take over, and the theory enters a superfluid phase. We can understand these holographic superconductor solutions as the endpoint of the instability towards condensing scalars found from a violation of the near-horizon AdS$_2$ BF bound (\ref{AdS2BF}).

One interesting question is how the condensate modifies the flow of charge. This can be studied in the conductivity, another transport coefficient similar to the shear viscosity discussed previously. This is captured in the spatial component of a frequency-dependent gauge field,
\begin{eqnarray}
\label{AConduct}
A_x(\omega) = A_x^{(0)}(\omega) + {A_x^{(1)}(\omega) \over r} + \ldots \,.
\end{eqnarray}
In practice this mode mixes with the metric fluctuation mode $h_{tx}$, and they must be studied together as a coupled system. 
From (\ref{AConduct}) one can then obtain the conductivity from Ohm's law $\langle \vec{J}\rangle =\sigma(\omega) \langle\vec{E}\rangle$, with $\langle E_x \rangle = \langle \dot{A}_x \rangle = -i \omega \langle A_x \rangle$, and thus
\begin{eqnarray}
\sigma(\omega) = {\langle J_x \rangle \over \langle E_x \rangle} = {i \over\omega} { A_x^{(1)}(\omega)\over A_x^{(0)}(\omega)} \,.
\end{eqnarray}
The $\omega \to 0$ limit gives the DC conductivity $\sigma\equiv \sigma(\omega=0)$ in a Kubo formula directly analogous to the one (\ref{Kubo}) for the shear viscosity. The result is plotted in figure~(\ref{fig:HoloSup}b) \cite{Hartnoll:2008vx} for a range of $T/T_c$; as $T$ decreases below the critical temperature, the low-frequency conductivity disappears, indicating an energy gap in the spectrum that eventually reaches size $\sim T$.

Top-down holographic superconductors embedded in string theory have also been found \cite{Gubser:2009qm, Gauntlett:2009dn, Gubser:2009gp, Gauntlett:2009bh, Ammon:2010pg}. There the large set of gauged fields and charged scalar fields leads to multiple instabilities and intricate webs of branches of superconducting solutions; for a description in the ABJM case, see \cite{Donos:2011ut}.

Zero-temperature limits of holographic superconductors are known to exist \cite{Horowitz:2009ij, Gubser:2009cg}, and can differ substantially from the extremal black holes we have already encountered as zero-temperature limits of non-superconducting systems.  One top-down zero-temperature solution (which can be viewed either as part of eleven-dimensional supergravity reduced to four dimensions \cite{Gauntlett:2009dn, Gubser:2009gp, Gauntlett:2009bh}, or directly in four-dimensional maximally supersymmetric gauged supergravity \cite{Bobev:2010ib}) takes the form of a {\em domain wall} geometry, where the spacetime is asymptotically anti-de Sitter both in the UV and IR ends. In the UV, the characteristic length scale is the usual one $L_{\rm UV}  = L$ associated to empty AdS space and the vacuum of ABJM theory, where the scalar fields vanish; in the IR, the running charged scalar approaches a different critical point of the supergravity potential, resulting in a different effective cosmological constant and a smaller AdS scale, $L_{\rm IR}$.  The rest of the geometry interpolates between these two AdS limits. Analogs of this geometry with no gauge field were found in \cite{Freedman:1999gp}, and can be thought of as RG flow geometries where a relevant perturbation in the UV brings the theory to a new nontrivial fixed point in the IR. In the zero-temperature superconductor case, the asymptotics of the scalar only lead to a field theory expectation value for the dual operator, but the chemical potential constitutes the deformation of the theory.

More generally, it is believed \cite{Gubser:2009cg} that zero-temperature holographic superconductors will interpolate between AdS in the UV, and a so-called Lifshitz geometry in the IR, with metric of the form \cite{Kachru:2008yh}
\begin{eqnarray}
ds^2_{\rm Lif} = -\left(r \over L_{\rm IR}\right)^{2z} dt^2 + {r^2 \over L_{\rm IR}^2} d\vec{x}^2 + {L_{\rm IR}^2 \over r^2} dr^2 \,,	
\end{eqnarray}
which has a scaling symmetry with exponent $z$ treating space and time differently,
\begin{eqnarray}
D: t \to \lambda t^z \,, \quad \vec{x} \to \lambda \vec{x}\,, \quad r \to {r \over \lambda} \,.
\end{eqnarray}

\section{Fermions and Strange Metals}

\subsection{High-temperature superconductors}

\begin{figure}
\begin{center}
\includegraphics[scale=0.28]{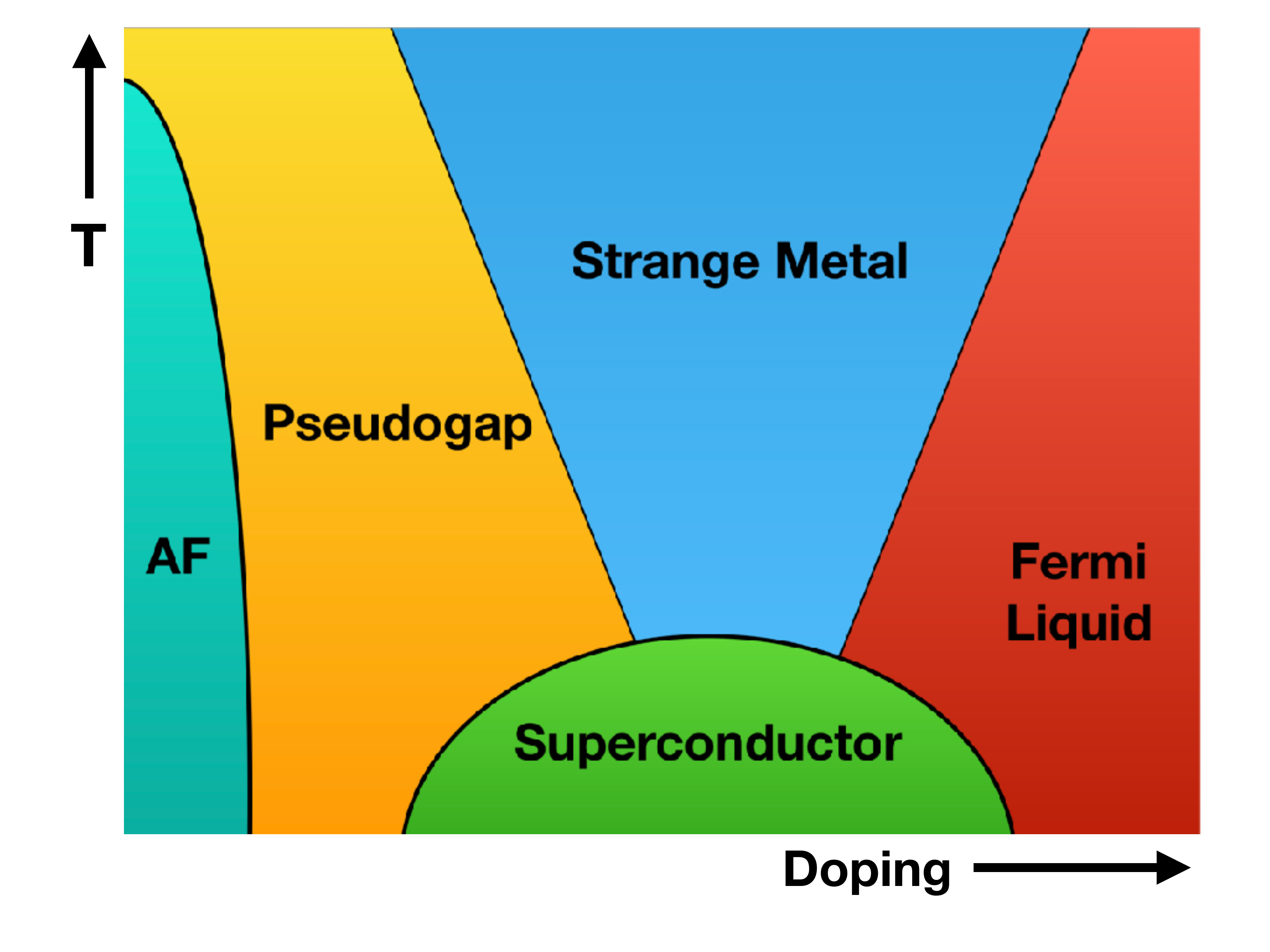}
\caption{A cartoon of the high-$T_c$ superconductor phase diagram, showing the superconducting dome, as well as Fermi liquid, strange metal, pseudogap and antiferromagnetic phases; the last two are not discussed here.
\label{fig:PhaseDiag2}}
\end{center}
\end{figure}

The so-called high-temperature (high-$T_c$) superconductors are interesting not just for their relatively high superconducting transition temperatures, but for a number of other interesting features of their phase diagrams, even outside the superconducting phase. A cartoon of the phase diagram is presented in figure~\ref{fig:PhaseDiag2}, where the vertical axis is the temperature, and the horizontal axis is a doping fraction of atoms in the lattice. Outside the superconducting region, at high doping the system acts like a traditional Fermi liquid: despite the interactions between electrons, it behaves as if transport is mediated by charged particles, which we can think of as electrons ``dressed" by the interactions. At zero temperature and finite density the dressed electrons arrange themselves into a Fermi surface, with quasiparticle excitations that are asymptotically stable as their energies approach the Fermi surface, as is familiar behavior from the usual Fermi theory of metals.

On the other hand, another phase exists above the superconducting dome, with more unusual properties. In this ``strange metal" phase, sharp Fermi surfaces exist, but there are no stable quasiparticles. Such a state of matter has been called a ``non-Fermi liquid" and remains theoretically challenging. From what we've learned about QCD in the quark-gluon plasma phase, it is tempting to think that a strongly-coupled phase of matter with no quasiparticle description might be well-described by a gravity dual. Can the gauge/gravity correspondence see a non-Fermi liquid?

To answer this question, we can look for Fermi surface singularities and the dispersion relations of their associated small fluctuations, by studying fermionic response in black hole gravity backgrounds. These backgrounds should have finite density, and the Fermi surface singularity should be cleanest at zero temperature (though finite temperature studies can be useful as well). To proceed, we will need to say something about AdS/CFT for fermions. For more discussion of this subject and its background, see for example \cite{Hartnoll:2016apf, McGreevy:2016myw, Iqbal:2011ae}.

\subsection{Fermions and AdS/CFT}

For fermionic excitations, the basic principle of AdS/CFT remains the same: each field on the gravity side contains boundary conditions corresponding to a source, while the response is free to fluctuate and is tied to the source through a boundary condition in the IR. However, because fermionic fields obey first order differential equations, things are encoded slightly differently: there is only one spinor's worth of boundary conditions from the higher-dimensional point of view, but this splits into  two separate spinors from the lower-dimensional perspective, one of which is the source and the other the dual operator expectation value \cite{Iqbal:2009fd}. As with scalar fields, for certain values of the mass there are two possible quantizations.

For a spin-1/2 field obeying the Dirac equation\footnote{We use conventions where $\{ \Gamma^\mu, \Gamma^\nu\} = - 2 g^{\mu\nu}$, with mostly plus signature metric. A hat indicates a flat-space index.}
\begin{eqnarray}
(i \Gamma^\mu \nabla_\mu - m) \chi = 0 \,,
\end{eqnarray}
the near-boundary solutions can be written in terms of the projections
\begin{eqnarray}
\chi_\pm \equiv {1 \over 2} \left( 1 \pm i \Gamma^{\hat{r}} \right) \chi\,,
\end{eqnarray}
as
\begin{eqnarray}
\chi_+ (r \to \infty) &=& A_+(t, \vec{x}) L^{d-2mL-1/2} \, r^{-d/2+mL} + \ldots\,, \\ \chi_- (r \to \infty) &=& A_-(t, \vec{x}) L^{d+2mL-1/2} \,  r^{d/2+mL} + \ldots \,,
\end{eqnarray}
As with the scalar example, we inserted factors of $L$ so that the engineering dimensions of the spinors $A_\pm$ match their $d$-dimensional scaling dimensions, $\Delta = d/2 \mp mL$.  Let us assume $m \geq 0$, otherwise we can effectively exchange $\chi_+$ and $\chi_-$. For $mL > 1/2$, we must take $A_+$ (the leading term) as the fixed spinor source:
\begin{eqnarray}
J_{\rm reg}(t, \vec{x}) =  A_+(t, \vec{x}) \,,
\end{eqnarray}
and allow $A_-$ to fluctuate, corresponding to the expectation value and leading to a dual operator of dimension $\Delta_\chi = {d \over 2} + mL$. For the window $0 \leq mL \leq 1/2$, the alternate quantization becomes possible as well, with source
\begin{eqnarray}
J_{\rm alt}(t, \vec{x}) =  A_-(t, \vec{x}) \,,
\end{eqnarray}
and dual operator dimension $\Delta_\chi = {d \over 2} + mL$. Thus the regular quantization with $A_+$ fixed gets us dimensions down to $\Delta = d/2$, and the alternate quantization fills in the window $(d-1)/2 \leq \Delta \leq d/2$, down to the unitarity bound.\footnote{Precisely at $m=0$, $\chi_+$ and $\chi_-$ have the same asymptotic scaling. Here both quantizations are possible, and they are equivalent in simple backgrounds but inequivalent in the presence of other interactions. This case is relevant for 4D ${\cal N}=8$ gauged supergravity, where SUSY can be used to select the regular quantization; see \cite{Breitenlohner:1982jf, Breitenlohner:1982bm} and for a modern discussion \cite{DeWolfe:2014ifa}.}

From the definition, each of $\chi_\pm$ contains half the degrees of freedom of $\chi$. When the gravity theory is in an even dimension, one can choose a basis for the $(d+1)$-dimensional Clifford algebra where $\Gamma^{\hat{r}}$ is diagonal and $A_\pm$ reduce to half-dimensional spinors appropriate for the odd-dimensional CFT$_d$; for example in AdS$_4$/CFT$_3$, a four-component Dirac spinor $\chi$ in 4D decomposes into two-dimensional Dirac spinors $\chi_\pm$ in 3D. Meanwhile, if the gravity theory is an odd dimension, $\Gamma^{\hat{r}}$ is proportional to the chirality matrix in the $d$-dimensional Clifford algebra, and the $\chi_\pm$ spinors are chiral; so in AdS$_5$/CFT$_4$, a four-component Dirac spinor in 5D becomes two Weyl spinors in 4D.

As with the bosonic case, in general we have to perform holographic renormalization for a fermion as well. Consider a Majorana fermion $\chi$ with $m=0$ in $d=3$, the case appropriate to 4D ${\cal N}=8$ gauged supergravity: here both the dual operator and its source have dimension $3/2$. We take the bulk + boundary action,
\begin{eqnarray}
S_{\rm Dirac} = {1 \over 2 \kappa^2}\int d^4x \sqrt{-g} {i \over 2} \bar\chi \Gamma^\mu \nabla_\mu \chi + {1 \over 4} \int d^3x \sqrt{-h} \, \bar\chi \chi  \,, 
\end{eqnarray}
which evaluated on solutions to the Dirac equation becomes,
\begin{eqnarray}
S_{\rm Dirac}  = {L^2 \over 4 \kappa^2} \int d^3x \,\bar{A}_+ A_- \,.
\end{eqnarray}
and leads to the action variation
\begin{eqnarray}
\delta S_{\rm Dirac} =  {L^2 \over 2 \kappa^2}\int d^3x \,   \bar{A}_- \delta A_+  \,,
\end{eqnarray}
which vanishes for the regular quantization $\delta A_+ = 0$, and leads to the  one-point function
\begin{eqnarray}
\label{FermiOnePoint}
\langle {\cal O}\rangle = {\delta S_{\rm Dirac} \over \delta \bar{J}} = {L^2 \over 2 \kappa^2 } A_- \,.
\end{eqnarray}
Changing the sign of the boundary term makes it suitable for the alternate quantization.

\subsection{Holographic Fermi and non-Fermi liquids}

Now that we understand fermionic correlation functions in AdS/CFT, we can
can search for Fermi surfaces in finite density systems by examining the fermionic Green's function
\begin{eqnarray}
G_R(k, \omega) \sim {A_- \over A_+} \,.	
\end{eqnarray}
(Since $A_\pm$ are spinors this is strictly speaking a matrix of Green's functions, but a choice of gamma matrices can diagonalize it.) One then defines a Fermi surface at momentum $k_F$ as a singularity in the Green's function at $\omega =0$, the energy of the Fermi surface:
\begin{eqnarray}
	G_R(\omega = 0, k = k_F) \to \infty \,.
\end{eqnarray}
Once a Fermi surface singularity is found, one can look for nearby fluctuations. For the extremal AdSRN geometries, there is a subtle order of limits issue between the near-horizon and small-$\omega$ expansions. Treating this with care relates the full Green's function $G_R$ near the Fermi surface to an auxiliary Green's function ${\cal G}(\omega)$ defined in the near-horizon $AdS_2$ region \cite{Faulkner:2009wj}. One then has the schematic form 
\begin{eqnarray}
G_R (\omega, k) \sim {1 \over k_\perp - {1 \over v_F} \omega + \ldots + {\cal G}(\omega)	}\,.
\end{eqnarray}
There is always a series in $\omega$, with higher order terms indicated by the ellipsis, but this may or may not be dominated by the IR Green's function
\begin{eqnarray}
{\cal G}(\omega) \sim \omega^{2 \nu_k} \,,	
\end{eqnarray}
where $\nu_k$ is an effective $AdS_2$ dual operator dimension. If $\nu_{k_F} > 1/2$, the leading small-fluctuation singularity is given by the $\omega/v_F$ term, which is real and hence is associated to an asymptotically stable mode: this is Fermi liquid behavior. However, if $\nu_{k_F}<1/2$, the dispersion relation is dominated by the complex ${\cal G}(\omega)$, which leads to unstable modes whose decay widths are of the same order as their energies; these are not asymptotically stable and describe a non-Fermi liquid.

In bottom up models in the AdSRN geometry, it was found that by tweaking the fermion mass and charge, both Fermi liquid and non-Fermi liquid behavior could manifest \cite{Lee:2008xf, Liu:2009dm, Cubrovic:2009ye, Faulkner:2009wj}. One can also study a set of top-down geometries dual to both ${\cal N}=4$ SYM and to ABJM theory; these geometries in general have running neutral scalars as well, but share the extremal near-horizon AdS$_2$ property of AdSRN. Instead of a doping parameter, one may vary the ratios of chemical potentials (of which one has three for the $SO(6)$ of ${\cal N}=4$, and four for the $SO(8)$ of ABJM) to produce new geometries. Holographic Fermi surfaces indeed appear in such top-down models for many fermionic fields, and interestingly, over a wide class of such top-down geometries, {\em every} fermion studied has non-Fermi liquid behavior \cite{DeWolfe:2011aa, DeWolfe:2012uv, DeWolfe:2014ifa}.\footnote{In certain special IR-singular geometries with vanishing entropy, there can be perfectly stable modes in an energy band, before the non-Fermi liquid behavior returns
\cite{DeWolfe:2013uba, DeWolfe:2014ifa}.}

Varying the chemical potentials can also produce geometries where $\nu_{k_F} \to 0$, which can be thought of as the divergence of a correlation length \cite{Iqbal:2011in}. This can indicate the boundary of a so-called ``oscillatory region", inside which no Fermi surface singularities exist as the Fermi momentum moves off into the complex plane; these regions are characterized by log $\omega$ terms and may be thought of existing in regions where the fermion charge is strong enough to allow pair production of charged particles in the IR AdS$_2$, in a fermionic version of the bosonic instability seen in holographic superconductors \cite{Faulkner:2009wj}. For certain fermions in top-down models there can also be isolated points where the correlation length diverges, so-called pole-zero transitions, where lines of Fermi surfaces (poles in $G_R$) transmute into zeros in $G_R$ instead.

\subsection{Fermionic response in holographic superconductors}

Besides the non-superconducting states just described, we can consider how fermions behave in holographic superconductors as well. Elementary superconducting states develop a mass gap: one question we can ask is, does such a gap occur for fermionic fluctuations in holographic superconductors?

\begin{figure}
\begin{center}
\includegraphics[scale=0.35]{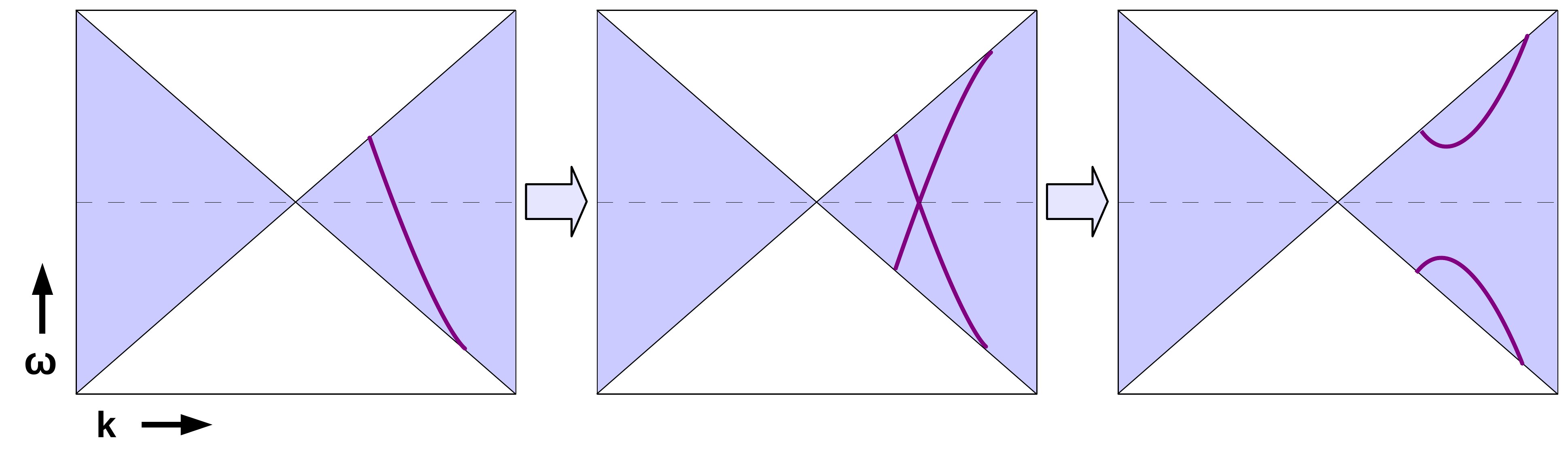}
\caption{A cartoon for the gapping mechanism of the Majorana BCS coupling. Without such a  coupling, fermionic excitations generally exist crossing the dashed Fermi surface, leading to an ungapped state. The antiparticles/holes have an identical line of excitation, flipped across $\omega=0$. Turning on the Majorana BCS coupling mixes these two energy bands, causing them to repel and leading to a gap. Taken from \cite{DeWolfe:2016rxk}.
\label{fig:LevelRepulsion}}
\end{center}
\end{figure}

Bottom-up fermions in superconducting backgrounds, with variable masses and charges,  generically display bands of ungapped stable fermionic modes, with higher charge in general leading to more bands \cite{Gubser:2009dt}. It was suggested by \cite{Faulkner:2009am} that such bands of excitations could be made gapped by a particular gravity interaction of the schematic form
\begin{eqnarray}
\label{MajoranaBCS}
S_{\rm Maj} = 	\int d^4x \, \phi \chi^T C \Gamma_5 \chi + \hbox{hermitian conjugate},  
\end{eqnarray}
where $\phi$ is the charged scalar active in the superconductor, and $\chi$ is the charged fermion; this Majorana-type coupling has net charge for the fermion bilinear, and can be thought of as an interaction between a Cooper pair $\chi \chi$ and the background condensate. Such an interaction is strongly reminiscent  of the BCS Hamiltonian for superconductivity, 
\begin{eqnarray}
H_{\rm BCS} = \Delta c^\dagger c^\dagger + 	\hbox{hermitian conjugate} \,,
\end{eqnarray}
with $c^\dagger$ a fermion creation operator and $\Delta$ the condensate.
In general, the excitations of particles and their antiparticles (holes)  have mirrored bands crossing at $\omega = 0$; the ``Majorana BCS" coupling (\ref{MajoranaBCS}) causes these bands to interact, and the resulting level repulsion pushes them away from the Fermi surface, leading to a gap (see figure~\ref{fig:LevelRepulsion}).

Thus in bottom-up models of holographic superconductivity, the ungapped state is generic, but a gap can be produced by suitably small charges or by an interaction of the form (\ref{MajoranaBCS}).
It is natural to ask what occurs in top-down systems. In one such zero-temperature model, an AdS$_4$ to AdS$_4$ domain wall describing a state in ABJM theory as described in the previous section, the fermionic excitations are gapped purely as a result of the charges being small \cite{DeWolfe:2016rxk}. In another, similar AdS$_4$ to AdS$_4$ domain wall state (technically not a superconductor since the scalar also has a source turned on, but geometrically very similar) \cite{Bobev:2011rv} the charges are large enough that without interactions an ungapped state would appear, but the theory provides a multi-field generalization of the Majorana BCS coupling (\ref{MajoranaBCS}), which conspires to precisely gap the fermionic modes \cite{DeWolfe:2015kma, DeWolfe:2016rxk}.

Thus, just as bottom-up models of non-superconducting finite density states could be Fermi liquids or non-Fermi liquids, but top-down systems appear to be exclusively non-Fermi liquids, bottom-up models of superconducting states can be gapped or ungapped, but top-down systems appear as exclusively gapped, for one reason or another. Whether this is conspiracy or coincidence is not entirely clear.

\subsection{Limits of the correspondence}

There is an interesting tension between these phenomena and what is visible in the bosonic variables.  In general bosonic response functions like the conductivity do not ``see" the corresponding Fermi surfaces; for example, a certain response to charged impurities called Friedel oscillations is expected to lead to singularities in the conductivity at $2k_F$, but these are not seen \cite{Blake:2014lva, Henriksson:2016gfm}. The oscillatory region and pole-zero transition behaviors, associated with something like a correlation length divergence and hence presenting the appearance of a quantum phase transition in the fermionic variables, do not show up in the thermodynamics or the bosonic response functions. It has been speculated that the fermionic response may be in some sense subleading in $N$ and hence invisible to the leading-$N$ bosonic response, and that leading order in $N$ requires a bulk condensation of fermionic fields, the so-called electron stars \cite{Hartnoll:2010gu,Hartnoll:2010ik}. However, the fermionic correlators appear at the same, leading order in $N$ as their bosonic counterparts; this is manifest in top-down models, where they show up at $N^2$ and $N^{3/2}$ respectively, essentially since they come from the same supergravity action as the bosons. This is apparently in conflict with the idea that the fermionic response is subleading in $N$.

This disconnect can be traced to the fact that two-point functions in the large-$N$ limit are calculated as linearized fluctuation equations on the gravity side, and so bosonic and fermionic gravity modes are ignorant of each other to this order; yet according to the deep reshuffling of degrees of freedom inherent in holography, they should each ``know" about both bosonic and fermionic degrees of freedom in the field theory. As the large N limit suppressed the quantum corrections to the scaling exponents in the holographic critical point, leaving mean field values, this mean field behavior of large N also seems to wash out the bosonic and fermionic responses' knowledge of each other.

We have seen that the AdS/CFT correspondence is a remarkable duality, and can be very useful for modeling the behavior of field theories at strong coupling without a quasiparticle description. We may not have an exact gravity dual for QCD or a laboratory system, but there are a number of lines of attack open to us: we can engineer a model holographic system designed to mimic the true one (as with the QCD phase diagram), we can look for universal behavior across a wide class of holographic models (as with the viscosity to entropy ratio), or we can try to find holographic models of broad classes of dynamics (as with superconductors and non-Fermi liquids). There are many other applications both already done and still yet to be done, and hopefully these lectures have provided a useful introduction to this broad and fascinating subject.

\acknowledgments

I am grateful to the organizers of TASI 2017, Mirjam Cveti\v{c} and Igor Klebanov, for inviting me to lecture and for organizing a wonderful school, and to Tom DeGrand and Emily Flanagan for running it so smoothly. I would also like to thank my fellow speakers for delivering a series of fascinating and enjoyable lectures. I would like to thank collaborators old and new who have taught me a lot about AdS/CFT, particularly Dan Freedman, Steve Gubser, Oscar Henriksson and Chris Rosen. Most of all I would like to thank the students, for their stimulating questions that made it such a pleasure to interact and talk about physics. My research is supported by the Department of Energy under Grant No.~DE-FG02-91-ER-40672. 

I would like to take a moment to remember Joe Polchinski. Joe had a long history with TASI, co-organizing TASI 1992 with Jeff Harvey where he delivered a now-classic set of lectures on effective field theory and the Fermi surface, speaking on D-branes at TASI 1996 right after they helped launch the second superstring revolution, giving a terrific set of talks on AdS/CFT at TASI 2010, and co-organizing the wonderful TASI 2015 summer school with Pedro Vieira, where he spoke on the black hole information problem. Without his discovery of D-branes, we would not have been led to the gauge/gravity correspondence as we were, and without his powerful thinking and mischievous sense of humor, the fields of quantum field theory, string theory and black hole information, among others, would have been much poorer. I learned of his death as these notes were being completed, and though they can only pretend to aspire to the level of clarity and insight that Joe would invariably bring to a subject, I would like to dedicate these lectures to him. 

\bibliographystyle{JHEP}
\bibliography{TASI_Lectures}

\end{document}